\documentclass[%
 reprint,
 superscriptaddress,
 amsmath,amssymb,
 aps,
]{revtex4-2}

\usepackage{graphicx}
\usepackage{dcolumn}
\usepackage{bm}

\usepackage{float}
\usepackage{amsmath}
\usepackage{amsfonts}

\usepackage[skip=2pt,font=footnotesize]{caption}
\usepackage[section]{placeins}
\usepackage{accents}

\begin{document}

\preprint{APS/123-QED}

\title{Multiple-Order Singularity Expansion Method}

\author{Isam Ben Soltane}
\email{isam.ben-soltane@fresnel.fr}
\affiliation{Aix Marseille Univ, CNRS, Centrale Marseille, Institut Fresnel, 13013 Marseille, France}
\author{Rémi Colom}
\affiliation{CNRS, CRHEA, Université Côte d’Azur, 06560 Valbonne, France}
\author{Félice Dierick}
\affiliation{Aix Marseille Univ, CNRS, Centrale Marseille, Institut Fresnel, 13013 Marseille, France}
\author{Brian Stout}%
\affiliation{Aix Marseille Univ, CNRS, Centrale Marseille, Institut Fresnel, 13013 Marseille, France}
\author{Nicolas Bonod}%
\email{nicolas.bonod@fresnel.fr}
\affiliation{Aix Marseille Univ, CNRS, Centrale Marseille, Institut Fresnel, 13013 Marseille, France}

\date{\today}

\begin{abstract}
Physical systems and signals are often characterized by complex functions of frequency in the harmonic-domain. The extension of such functions to the complex frequency plane has been a topic of growing interest as it was shown that specific complex frequencies could be used to describe both ordinary and exceptional physical properties. In particular, expansions and factorized forms of the harmonic-domain functions in terms of their poles and zeros under multiple physical considerations have been used. In this work, we start from a general property of continuity and differentiability of the complex functions to derive the multiple-order singularity expansion method. We rigorously derive the common singularity and zero expansion and factorization expressions, and generalize them to the case of singularities of arbitrary order, whilst deducing the behaviour of these complex frequencies from the simple hypothesis that we are dealing with physically realistic signals.
\end{abstract}

\maketitle


\section{Introduction}
\begin{figure*}
    \centering
    \includegraphics[width=0.85\textwidth]{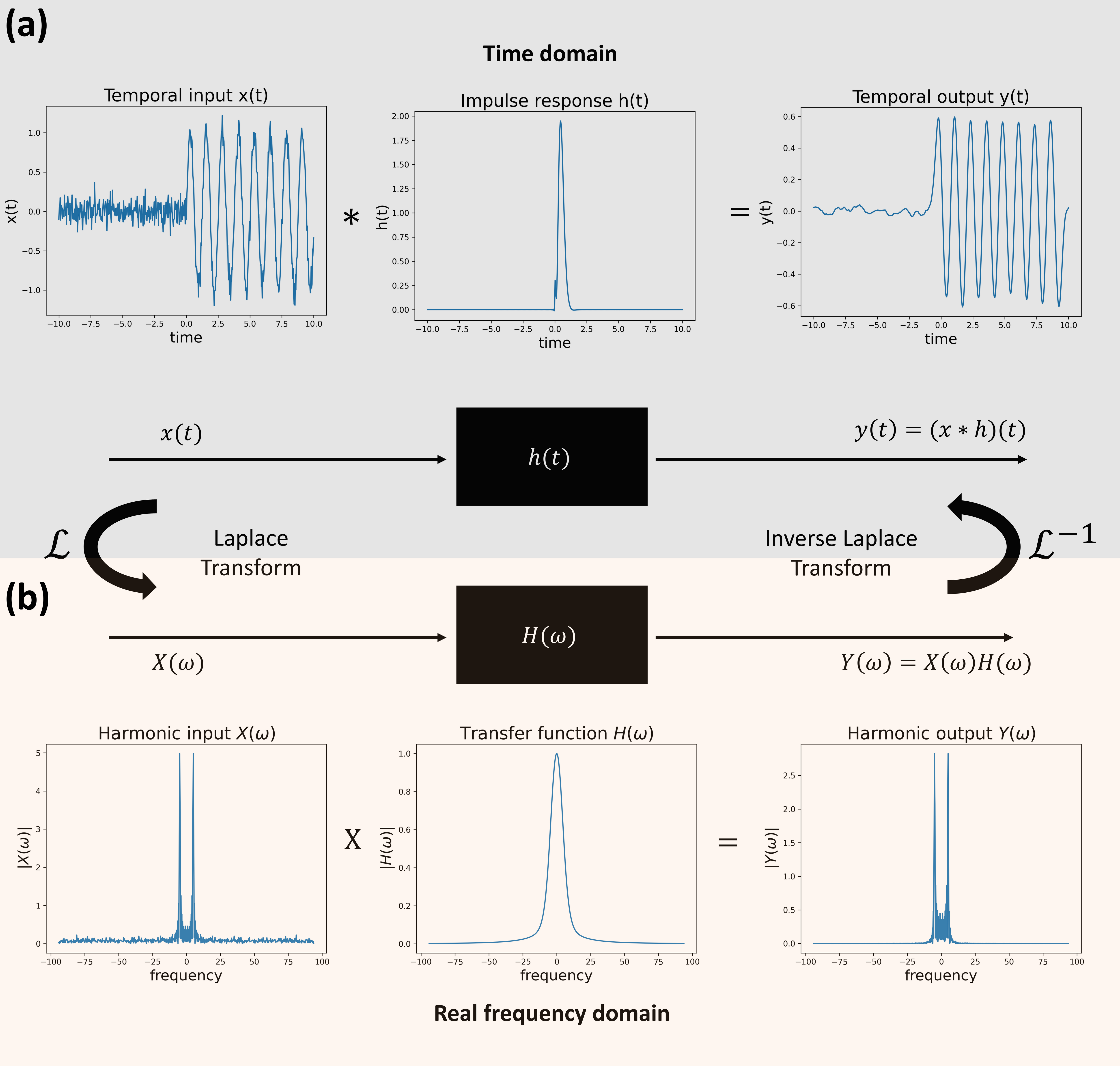}
    \caption{General representation of an LTIS. (a) In the temporal domain, the system is characterized by its impulse response $h(t)$ which can be used to obtain the output $y(t)$ after a convolution with the input $x(t)$. (b) In the harmonic domain, the LTIS is described by its transfer function $H(\omega)$. It is a filter acting in the frequency domain on the input $X(\omega)$ to generate an output $Y(\omega)$. In this case, $H(\omega)$ is a low-pass filter which partially removes the noise present at higher frequencies, as it can be seen by comparing the temporal input $x(t)$ to the temporal output $y(t)$. The output $Y(\omega)$ is the product of $H(\omega)$ and $X(\omega)$. The harmonic domain functions $X(\omega)$, $H(\omega)$ and $Y(\omega)$ are obtained by bilateral Laplace transform of the temporal signals $x(t)$, $h(t)$ and $y(t)$. The temporal signals can then be recovered using an inverse Laplace transform.}
    \label{fig:LTIS_viz}
\end{figure*}

The model of Linear and Time-Invariant Systems (LTIS) is commonly used in physics to derive the response of a medium to a excitations. Such systems are usually studied in the harmonic domain where they are associated with complex-valued transfer functions that can be used to fully describe the response of the system to an arbitrary excitation~\cite{dazzo1983,oppenheim1997}. This formalism has been mainly developed in the research field of automatic system control but it can be found in a wide variety of problems where it goes by different names. The complex impedance and admittance formalism, for instance, is used to characterize the properties of materials in electronics, acoustics and biology~\cite{beranek1942,ackmann1993,callegaro2012}. The Scattering matrix formalism, first introduced in quantum electrodynamics~\cite{dyson1949}, can be used to link outgoing waves to incoming waves, or final states to initial states in scattering problems in various situations (transport phenomena, diffraction gratings, integrated circuits, chaotic systems, ...)~\cite{agassi1975,popov1986,leijtens1996,fyodorov2010}.

While emphasis is generally placed on real and positive frequencies with clear physical interpretations, the description of a linear transfer system together with its input and response signals often requires an analytical continuation into the complex frequency plane~\cite{williams2002,krasnok2019}. Instead of considering individual responses at specific real frequencies, complex frequencies can provide us with intuition as to how a system will behave or how a signal is shaped over a large spectral width. The analysis of the electromagnetic response in terms of complex zeros and singularities has turned out to be highly efficient for several applications such as quantum waveguides~\cite{porod1993}, highly selective filters~\cite{tsuzuki2002}, plasmonic metasurfaces~\cite{grigoriev2013_fano}, anapoles~\cite{colom2019modal}, coherent perfect optical absorbers~\cite{grigoriev2015,baranov2017,zhan2014,wang2021,chen2022use} and analog computing~\cite{sol2022}. In addition, studying the order of the zeros and singularities can help in providing better interpretations of the associated phenomena~\cite{desoer1974}.

Any real temporal signal $h(t)$ can be associated with a harmonic-domain function $H(\omega)$, which is either an input or output signal, or a transfer function linking the two and describing an LTIS as a filter acting on the input in the harmonic domain (see Fig.~\ref{fig:LTIS_viz}). Different methods have been developed to express $H$ with respect to its zeros and singularities in the complex frequency plane. The most relevant methods, in our case, can be traced back to the Weierstrass and Hadamard factorization theorems for holomorphic and meromorphic functions respectively, which provide general expressions of $H$ as opposed to local expressions such as the Laurent series expansion. 

$(i)$-With the rapid development of electronics and automated machines, the control of the LTIS stability often favoured the factorization of the transfer function $H$, in the simple form of the ratio of complex polynomials~\cite{sanathanan1963}, in order to monitor the evolution of the phase and amplitude with respect to the frequency in Bode diagrams~\cite{koenig1959}. In electronics and automatic system control, singularities and zeros of multiple-order are not uncommon ($n^{\rm th}$ order Butterworth filters~\cite{butterworth1930} for instance). 

$(ii)$ Alongside this progress, pioneer works led to the development of the Singularity Expansion Method (SEM)~\cite{baum1971,baum1986,baum2012}. It was first developed to approximate the time dynamics of systems associated with arbitrary-order singularities in the harmonic-domain and helped in describing many problems in electromagnetism~\cite{vincent1978,vincent1979,vincent1989,baum1991,baum2012} (an exhaustive set of references dating back to the early developments of the SEM can be found in ref.~\cite{michalski1981}). The SEM has received a renewed interest in the recent years in the case of simple poles, \textit{i.e.} singularities of order 1, where accurate expressions have been derived and applied to various problems~\cite{arfken2005,grigoriev2013,mansuripur2017,colom2018,bensoltane2022,colom2022topology}. In addition, pole or singularity expansions have been increasingly used in the frame of quasi-normal modes or resonant state expansions~\cite{defrance2020,benzaouia2021,sauvan2021,sauvan2022,ammari2022}. However, exact expansions have not been obtained in the case of poles of arbitrary order, despite the prospects they offer in describing more complex systems and/or their input and output signals ~\cite{miri2019,sweeney2019,ermolaev2022}.

These two methods, \textit{i.e.} the singularity expansion method for simple poles and the multiple-order pole and zero factorization used in system control and electronics, can be obtained under certain restrictive considerations. Furthermore, the derivation of one method from the other and their relationships are not obvious, in spite of the same singularities appearing in both the factorized and expanded expressions.

In this work, we first describe and explain the usual state of the art expressions of the pole expansion in the case of simple poles, and the pole and zero factorization as depicted in the fields of electronics and automatic system control. We present some of the limits of these expressions which motivated the need for the general expressions of the singularity expansion and the singularity and zero factorization which are then rigorously derived using complex analysis theorems applied to the harmonic-domain signals at play under simple hypothesis. We show that the behaviour of any function around its discrete set of singularities and/or zeros is enough to describe that function in the complete complex frequency plane. In addition, we put the emphasis on the constraints of the distribution of the singularities and zeros which arise from the general, non-restrictive consideration of physically realistic temporal signals, \textit{i.e.} real and causal signals. Finally, we show how to obtain, from these harmonic-domain expressions, the temporal singularity expansion which holds information regarding the stability of the systems, the convergence of the signals, and the transient and steady-state temporal dynamics.

\section{Expansion and Factorization Using Poles and Zeros}\label{sec:harmonic_exp}

\subsection{Simple Pole Expansion}

\begin{figure*}
    \centering
    \includegraphics[width=0.9\textwidth]{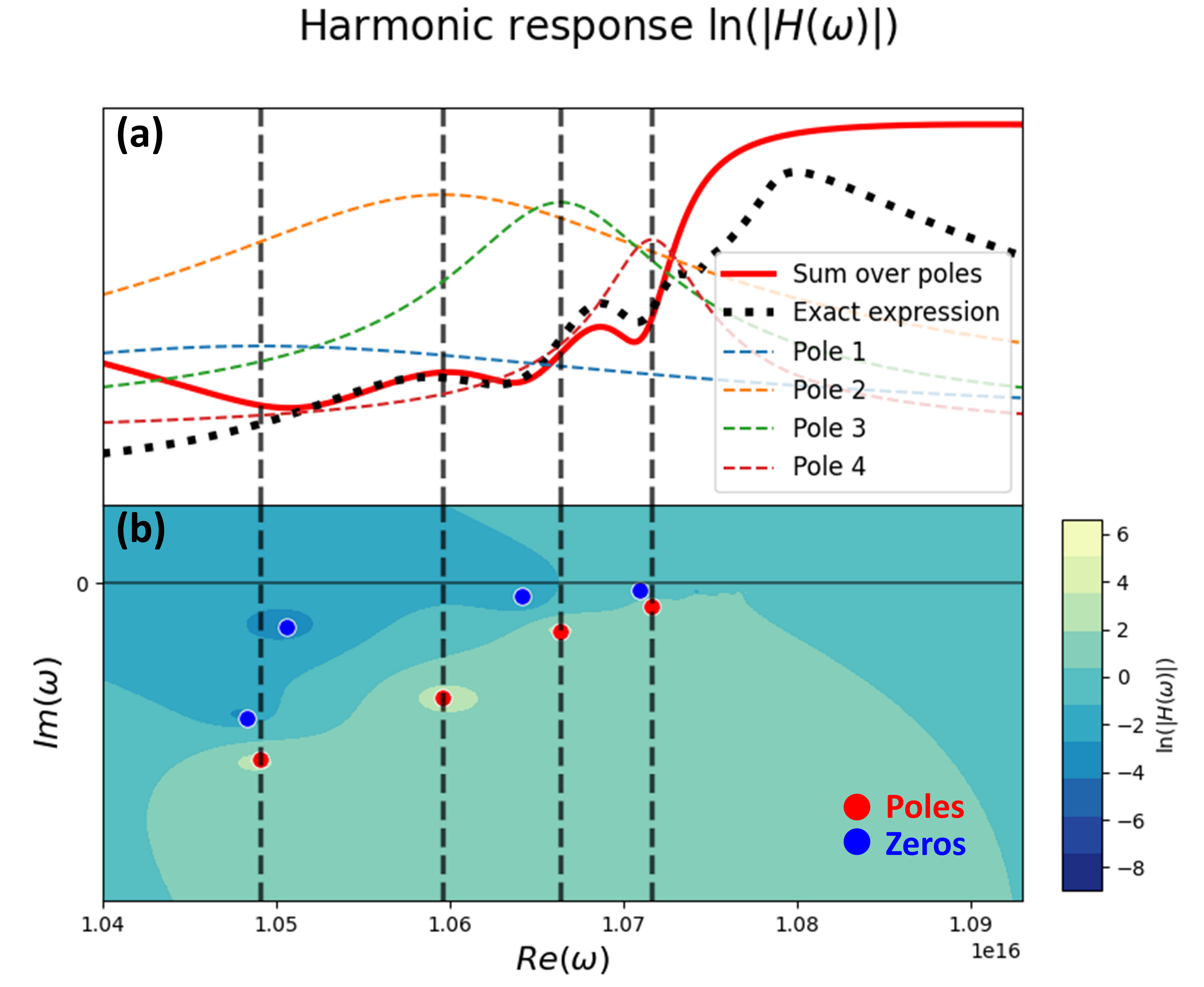}
    \caption{(a) Modulus of the transfer function $H(\omega)$ defined in Eq.~(\ref{eq:H_example}), as well as its resonant terms $Res(p_\ell)/(\omega - p_\ell)$ associated with the poles $p_l$ and the modulus of the sum of these resonant terms. The frequencies range from $1.040 \times 10^16$~Hz to $1.093 \times 10^16$~Hz, which is equivalent to ultraviolet wavelengths between $172.9$~nm and $181.3$~nm. (b) Log-amplitude of $H(\omega)$ in the complex frequency plane, in a limited complex frequency window. The poles in (a) are highlighted in (b) as red points. The sum over the poles gives an approximation of the shape of $H(\omega)$ on the real axis (the minimum and maximum frequencies), but with the background term of Eq.~(\ref{eq:spe}) omitted, the reconstructed red curve poorly matches the exact expression.}
    \label{fig:fig2_shape_bgr}
\end{figure*}

The first singularity expansions were developed in the aforementioned SEM~\cite{baum1971,baum1986,baum2012} based on the observation of a system's response to a sinusoidal input in the transient regime. It was shown to be a combination of damped sinusoidal functions associated with complex frequencies which were the singularities of the transfer function of the system. While Baum first derived the singularity expansion by taking into account the order of these singularities, most applications made use of (and later demonstrated) this expansion in the case of simple poles, \textit{i.e} isolated singularities of order 1~\cite{arfken2005,grigoriev2013,colom2018}. The resulting expansion is referred to as the Simple Pole Expansion (SPE), and it has the following expression:
\begin{eqnarray}
    \begin{aligned}
        &H(\omega) = H_{\text{bgrd}}(\omega) + \sum_{p} \frac{\text{Res}(H,p)}{w-p} \\
    \end{aligned}
    \label{eq:spe}
\end{eqnarray}
where $p$ denotes the simple poles of $H$. 
The SPE is composed of a background term $H_{\text{bgrd}}(\omega)$ which does not possess poles~\cite{benzaouia2021}, and a sum of resonant terms $\text{Res}(H,p)/(\omega - p)$. This translates the idea that the shape of $H(\omega$) in a specific frequency range is chiefly affected by the nearby singularities, with the background term compensating for the offset introduced by the singularities with respect to the average (or background) value of $H$ in that range. When all the poles $p$ of $H$ are known, the background term reduces to a constant. The contribution of the poles to the shape of $H$ is illustrated in Fig.~\ref{fig:fig2_shape_bgr} for $H$ defined as the reflection coefficient of a thin layer of silver illuminated from one side at normal incidence:
\begin{eqnarray}
    \begin{aligned}
        H(\omega) = r(\omega) - \frac{t(\omega)t'(\omega)r(\omega)e^{2i\omega \frac{n(\omega)d}{c}}}{1 - r(\omega)^2e^{2i\omega \frac{n(\omega)d}{c}}}
    \end{aligned}
    \label{eq:H_example}
\end{eqnarray}
where $d=70$~nm is the thickness of the silver layer, $c$ is the speed of light in the air, $n(\omega)$ is the refractive index of silver (see the SI for the detailed expression), $r(\omega)=(n(\omega)-1)/(n(\omega)-1)$ is the Fresnel reflection coefficient at the air/silver interface, and $t(\omega)=2/(n(\omega)+1)$ and $t'(\omega)=2n(\omega)/(n(\omega)+1)$ are the Fresnel transmission coefficients at the air/silver and silver/air interfaces respectively. The zeros, the poles and the residues of $H$ were determined numerically. Each pole $p$ in the complex frequency plane, in Fig.~\ref{fig:fig2_shape_bgr} (b) is associated with a resonant term $\text{Res}(H,p)(\omega - p)$, whose  moduli are plotted in Fig.~\ref{fig:fig2_shape_bgr} (a). The sum of the resonant terms associated with the poles in the plotted complex frequency window gives the red curve in Fig.~\ref{fig:fig2_shape_bgr} (a), which matches the local shape of the exact response (dashed line which corresponds to Eq.~(\ref{eq:H_example}), \textit{i.e.} the approximate position of the local minimum and maximum frequencies. The poor match between this sum of resonant terms and the exact expression is due to the missing background term $H_{\text{bgrd}}(\omega)$ which should correct for the influence of the poles outside of that frequency window. 

Since most physical systems are described by poles of order 1, the SPE is well suited for the description of the transfer function of such systems, or their response to a temporal or spatial sinusoidal input~\cite{grigoriev2011,grigoriev2013,GarciaVergara2017,colom2018,bensoltane2022}. Other expressions must be used when the order of the poles is increased. One commonly used alternative in electrical and electrics engineering is what we will refer to as the Pole and Zero Factorization (PZF).

\subsection{Pole and Zero Factorization}

The PZF is obtained by making the assumption that the harmonic-domain function $H$ is the ratio of two polynomials of the complex frequency variable $\omega$~\cite{sanathanan1963,dazzo1983,oppenheim1997}, and that any system responds instantaneously to an input signal:
\begin{eqnarray}
    \begin{aligned}
        H(\omega) = \frac{N(i\omega)}{D(i\omega)} = \frac{N_0\prod_{\ell=1}^{deg(N)}(i\omega - iz_\ell)}{D_0\prod_{\ell=1}^{deg(D)}(i\omega - ip_\ell)}
    \end{aligned}
    \label{eq:PZF}
\end{eqnarray}
where $iz_\ell$ and $ip_\ell$ are the zeros of $N$ and $D$ respectively, with potentially identical zeros and poles (in which case their order is higher). Let us point out that we consider, in this case, the variable $i\omega$ instead of $\omega$ for $N$ and $D$ in order to remain consistent with the usual definition of the Laplace transform formalism. Using a partial fraction expansion, it is possible to write an expression similar to the SPE:
\begin{eqnarray}
    \begin{aligned}
        H(\omega) &= \alpha_{1,0} + \frac{\alpha_{1,1}}{\omega - p_1} + ... + \frac{\alpha_{1,2}}{(\omega - p_1)^{\nu_1}} \\
        &+ \alpha_{2,0} + \frac{\alpha_{2,1}}{\omega - p_2} + ... + \frac{\alpha_{2,2}}{(\omega - p_2)^{\nu_2}}. \\
        &+ ... \\
        &= H_{\text{bgrd}} + \sum_{\ell} \sum_{m=1}^{\nu_\ell} \frac{\alpha_{\ell,m}}{(\omega - p_\ell)^{m}}
    \end{aligned}
\end{eqnarray}
where $H_{\text{bgrd}} = \sum_\ell \alpha_{\ell,0}$ and the poles $p_\ell$ are considered with their order or multiplicity $\nu_\ell$.  The constants $\alpha_{\ell,1}$ can be identified as the residues, and they can be calculated with the zeros and poles using Eq.~(\ref{eq:PZF})~\cite{baum1971,grigoriev2013}:
\begin{eqnarray}
    \begin{aligned}
        \alpha_{k,1} &= \text{Res}(H, p_k) \\
        &= \lim_{\omega \rightarrow p_k} (\omega - p_k)H(\omega) \\
        & = -i\frac{N_0\prod_{\ell=1}^{deg(N)}(ip_k - iz_\ell)}{D_0\prod_{\ell \neq k}^{deg(D)}(ip_k - ip_\ell)}
    \end{aligned}
\end{eqnarray}

As stated in the introduction, the PZF and the resulting pole expansion are widely used in electronics and system control where they provide highly accurate results when studying systems at time scales larger than the transient time scale. However, this is not the case for other fields such as wave physics, in which case the PZF is missing a complex exponential factor depending on the frequency $\omega$ which can be interpreted as the result of a time delay required by a system to respond to different frequencies (\textit{i.e. a dispersive system)}~\cite{grigoriev2013,colom2018}. This term, which is particularly important for systems which interact with signals with respect to both time and space cannot be obtained from the restrictive hypothesis that $H$ is a ratio of two polynomials. 
In the next section, we present a singularity expansion taking into account the potentially infinite order of the singularities. From this expression, a more general singularity and zero factorization will then be derived.

\section{Multiple-Order Singularity Expansion and Factorization} 
\label{sec:multpli_harmonic_exp}
\subsection{Multiple-Order Singularity Expansion}

We now consider a meromorphic function $H$, \textit{i.e.} holomorphic everywhere on $\mathbb{C}$ except for a set of points $\mathcal{P}$ which is the set of its singularities which are all assumed to be poles or isolated essential singularities.
Using the Cauchy integration theorem and the residue theorem, the function $H$ can be expressed as the sum of an integral term and an expansion on its set of singularities (see Eqs.~(S1 - S5) in the SI):
\begin{eqnarray}
    \begin{aligned}
        H(\omega) &= \frac{1}{2i\pi}\int_{\gamma} \frac{H(z)}{z-\omega} dz - \sum_{p} \text{Res}(F,p) \\
        F(z) &= \frac{H(z)}{z-\omega}
    \end{aligned}
\end{eqnarray}
where $\gamma$ is a closed curve around the singularities $\{p\}$ of $H$, and $\text{Res}(F, p)$ is the residue of $F$ at the singularity $p$. The residues of $F$ can be analytically determined from the Laurent series coefficients $\alpha(H, p, n)$ of $H$ around the poles $p$, which provides local information regarding the behaviour of $H$ in the vicinity of its singularities. This leads to the following expression (details of the calculations can be found in Eqs.~(S6 - S17) of the SI):
\begin{eqnarray}
    \label{eq:f_ex}
    H(\omega) = \frac{1}{2i\pi}\int_{\gamma} \frac{H(z)}{z-\omega} dz + \sum_{p} \sum_{m=1}^{\nu_p} \frac{\alpha(H, p, -m)}{(\omega-p)^m}
\end{eqnarray}
where $\nu_p$ is the potentially infinite order of the singularity $p$. Eq.~\ref{eq:f_ex} shows two contributions to the behaviour of $H$ at the frequency $\omega$: (i) the nearby singularities $\{p\}$, within the closed curve $\gamma$; (ii) the set of values of $H$ on $\gamma$, represented by the integral term, which accounts for the contribution of all the singularities outside of $\gamma$. The bigger the closed curve gets, the less the integral affects the value of $H$ since the singularities outside the closed curve become too far from $\omega$, provided that $H$ does not grow faster than $|\omega|$ in the complex plane (which is usually the case for physically realistic signals). When this hypothesis holds, we show that the integral term can be replaced by a constant by replacing the curve by a circle of infinite radius (Eqs.(S18-S24) of the SI):
\begin{eqnarray}
    \begin{aligned}
        &H(\omega) = H_{\text{NR}} + H_{\text{R}}(\omega)
    \end{aligned}
    \label{eq:f_approx}
\end{eqnarray}
where $H_{\text{NR}}$ is a constant non-resonant term, and $H_{\text{R}}(\omega)$ is a resonant shaping term which depends on the frequency $\omega$. The non-resonant term $H_{\text{NR}}$ is defined as:
\begin{eqnarray}
    \begin{aligned}
        &H_{\text{NR}} = H(a) - \sum_{p} \sum_{m=1}^{\nu_p} \frac{\alpha(H, p, -m)}{(a-p)^m}
    \end{aligned}
    \label{eq:f_approx_Hres}
\end{eqnarray}
where $a$ an arbitrary complex frequency which is not a pole. Let us stress that the choice of $a$ has no influence over the value of the constant term $H_{\text{NR}}$. 
The resonant term $H_{\text{R}}(\omega)$ possesses the following expression:
\begin{eqnarray}
    \begin{aligned}
        &H_{\text{R}}(\omega) = \sum_{p} \sum_{m=1}^{\nu_p} \frac{\alpha(H, p, -m)}{(w-p)^m}
    \end{aligned}
    \label{eq:f_approx_Hshape}
\end{eqnarray}
If $0$ is not a pole, we usually set $a=0$ and the non-resonant term $H_{\text{NR}}$ can be expressed using the static response $H(0)$:
\begin{eqnarray}
    \begin{aligned}
        &H_{\text{NR}} = H(0) - \sum_{p} \sum_{m=1}^{\nu_p} \frac{\alpha(H, p, -m)}{(-p)^m}
    \end{aligned}
    \label{eq:f_approx_res}
\end{eqnarray}
We refer to Eqs.~(\ref{eq:f_approx}) to~(\ref{eq:f_approx_Hres}) as the Multiple-Order Singularity Expansion Method (MOSEM). The accuracy of MOSEM is shown in the case of poles of order 2 in a purely theoretical example in the SI. When the order $\nu_p$ of all the poles is 1, the expression becomes:
\begin{eqnarray}
    \begin{aligned}
        H(\omega) &= H_{\text{NR}} + \sum_{p} \frac{\alpha(H,p,-1)}{\omega - p}
    \end{aligned}
\end{eqnarray}
Let us stress that the residue of $H$ associated with the pole $p$ is defined as $\text{Res}(H,p)=\alpha(H,p,-1)$. Therefore by identifying $H_{\text{NR}}$ as a constant background term $H_{\text{bgrd}}$ when all the singularities are taken into account in the resonant terms, we obtain the SPE expression defined in Eq.~(\ref{eq:spe}):
\begin{eqnarray}
    \begin{aligned}
        H(\omega) &= H_{\text{bgrd}} + \sum_{p} \frac{\text{Res}(H,p)}{\omega - p}
    \end{aligned}
\end{eqnarray}
If $H$ is reconstructed using only a finite set of singularities, two strategies can be adopted to still obtain a good match between the resulting truncated SPE or MOSEM and the exact expression; (i) as mentioned before, $H_{\text{bgrd}}$ can be considered as a holomorphic function, \textit{i.e.} with no singularities, which models the influence of the singularities missing from the resonant term $H_{\text{R}}(\omega)$; (ii) the influence of the poles outside of the region of interest can be accounted for by considering an additional virtual singularity within the resonant term as was done in ref~\cite{bensoltane2022}.

\subsection{Singularity and Zero Factorization}

\begin{figure*}
    \centering
    \includegraphics[width=\textwidth]{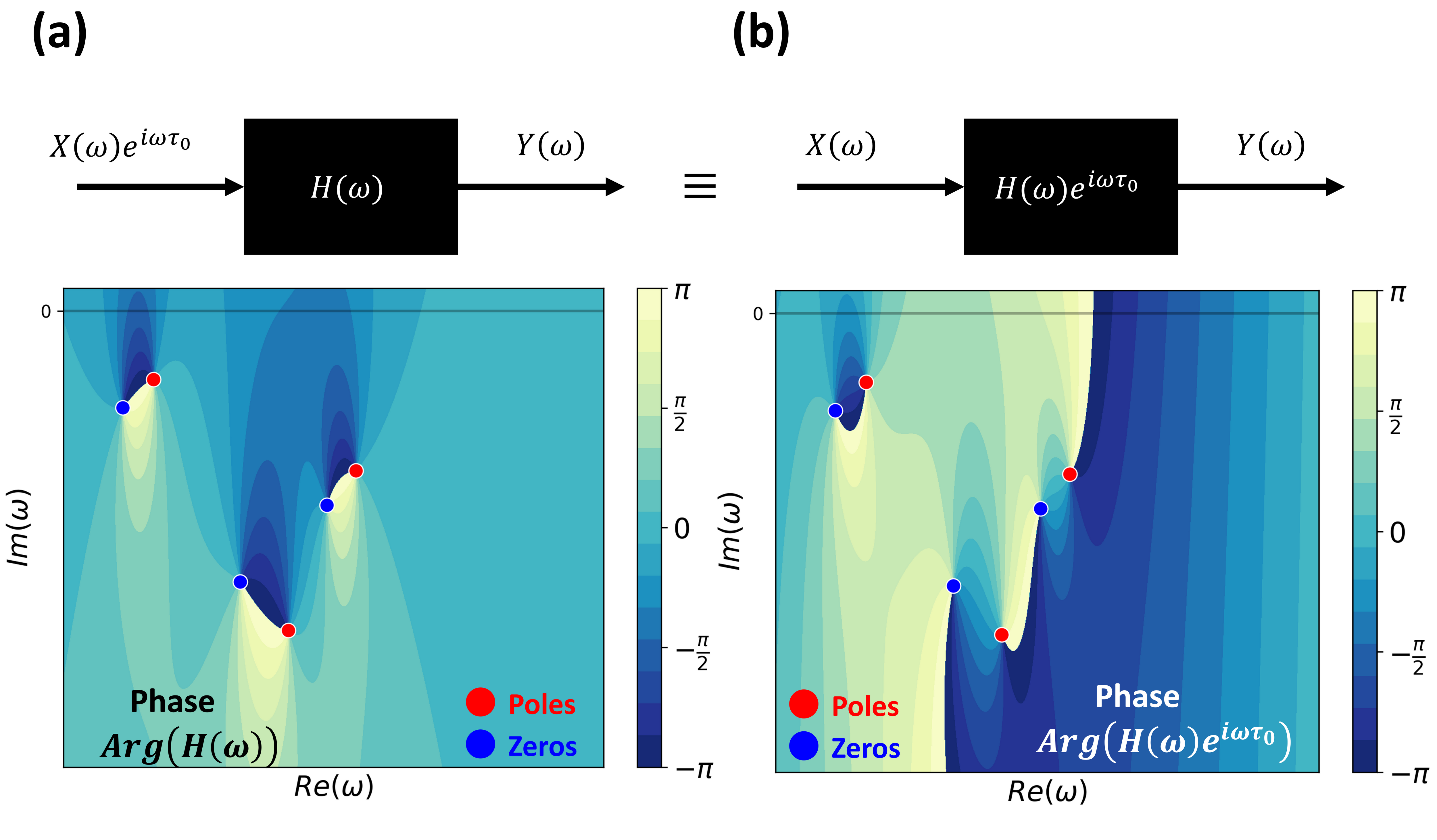}
    \caption{(a) Phase diagram in the complex frequency plane of the transfer function $H(\omega)$ of an LTIS defined in Eq.~(\ref{eq:H_example_2}) as $H(\omega) = 5 + \frac{2 + 0.1i}{\omega - (2 + 0.3i)} + \frac{4 + 0.1i}{\omega - (5 + 1.4i)} + \frac{6 + 0.1i}{\omega - (6.5 + 0.7i)}$. (b) Phase diagram of the transfer function $H(\omega)e^{i\omega\tau_0}$, with $\tau_0=\pi/6$, of the same LTIS with a distinct time origin. Shifting the time-origin of the input $X(\omega)$ in (a) is tantamount to considering $H(\omega)e^{i\omega\tau_0}$ for the transfer function in (b). The distribution of the poles (red points) and zeros (blue points) remains the same, only the phase is affected.}
    \label{fig:fig3_to}
\end{figure*}

The Weierstrass and Hadamard factorization theorems state that any meromorphic function can be written as the ratio of two complex polynomials multiplied by a complex exponential. Starting from this consideration, expressions such as the PZF (in which the complex exponential is usually removed) can be obtained (although not straightforwardly) to study the phase and amplitude variations at real frequencies. The poles and zeros can be defined in the resulting rational fractions as the zeros of the denominator and numerator respectively. Alternatively, a factorization involving the singularities and zeros can be obtained from MOSEM, providing a clearer link between the expanded and factorized forms as was done in Refs.~\cite{grigoriev2013,grigoriev2013_fano} in the case of poles of order 1.

Here, we aim at deriving a generalized factorized expression from MOSEM expression of the function $H$ in Eqs.~(\ref{eq:f_approx}) to~(\ref{eq:f_approx_Hres}). $H$ can always be written as:
\begin{eqnarray}
    \begin{aligned}
        H(\omega) = \omega^m G(\omega)
    \end{aligned}
\end{eqnarray}
with $m\geq0$ the order of the zero $0$ of $H$, and $G$ a meromorphic function which does not possess $0$ as a zero. If $m=0$, we have $G=H$.
Let us consider $F$ the log-derivative of $G$, with $G'$ the complex derivative of $G$:
\begin{eqnarray}
    \begin{aligned}
         &F = \frac{G'}{G} \\
         &G' = \partial_\omega[G]
    \end{aligned}
\end{eqnarray}
$G'$ is calculated by taking the derivative of MOSEM applied to $G$ (Eqs.~(\ref{eq:f_approx}) to~(\ref{eq:f_approx_Hres})). The poles of $F$ are the zeros $z_\ell$ and the singularities $p_\ell$ of $G$, and they are all of order 1 ((see Eqs.~(S33 - S36) in the SI): 
\begin{eqnarray}
    \begin{aligned}
        &\forall z_\ell \in \mathbb{C}, G(z_\ell)=0 \iff F(z)=\infty \\
        &\forall p_\ell \in \mathbb{C}, G(p_\ell)=\infty \iff F(z)=\infty \\
    \end{aligned}
\end{eqnarray}
In addition, their associated residues are the order $\nu_z$ and $\nu_p$ of the zeros and poles of $G$ respectively :
\begin{eqnarray}
    \begin{aligned}
        &\forall (z_\ell, \nu_z), ~ \text{Res}(F, z) = +\nu_z \\
        &\forall (p_\ell, \nu_p), ~ \text{Res}(F, p) = -\nu_p
    \end{aligned}
\end{eqnarray}
We apply MOSEM to $F$ taking these values into account:
\begin{eqnarray}
    \begin{aligned}
        &F(\omega) := F_{\text{NR}} + F_{\text{R}}(\omega) \\
        &F_{\text{NR}} := F(a) - \sum_{p} \sum_{m=1}^{\nu_p} \frac{\alpha(F, p, -m)}{(a-p)^m} \\
        &F_{\text{R}}(\omega) := \sum_{p} \sum_{m=1}^{\nu_p} \frac{\alpha(F, p, -m)}{(w-p)^m}
    \end{aligned}
\end{eqnarray}
with $p\in\{z_\ell, p_\ell\}$, $\nu_p=1$, and $\alpha(F, p, -1) = 1$ if $p=z_\ell$ is a zero of $H$ and $\alpha(F, p, -1) = -1$ if $p=p_\ell$ is a pole of $H$:
\begin{eqnarray}
    \begin{aligned}
        &F(\omega) = F_{\text{NR}} + F_{\text{R}}(\omega) \\
        &F_{\text{NR}} = F(a) + \sum_{p_\ell} \frac{\nu_{p_\ell}}{a-p_\ell} - \sum_{z_\ell} \frac{\nu_{z_\ell}}{a-z_\ell} \\
        &F_{\text{R}}(\omega) = -\sum_{p_\ell} \frac{\nu_{p_\ell}}{\omega-p_\ell} + \sum_{z_\ell} \frac{\nu_{z_\ell}}{\omega-z_\ell}
    \end{aligned}
\end{eqnarray} 
Finally, we obtain the following expression of $F$:
\begin{eqnarray}
    \begin{aligned}
        F(\omega) = \frac{G'(a)}{G(a)} &+\sum_{z_\ell} \left( \frac{\nu_{z_\ell}}{\omega-z_\ell} - \frac{\nu_{z_\ell}}{a-z_\ell} \right) \\
        &- \sum_{p_\ell} \left( \frac{\nu_{p_\ell}}{\omega-p_\ell} - \frac{\nu_{p_\ell}}{a-p_\ell} \right)
    \end{aligned}
    \label{eq:Log_deriv_simple}
\end{eqnarray}
By integrating $F$ on a curve from the arbitrary complex frequency $a$ to the frequency of interest $\omega$, and applying the exponential function to the result, we derive the following expression (Eqs.~(S37 - S41) in the SI):
\begin{eqnarray}
    \begin{aligned}
        &H(\omega) = \omega^m ~ G(a) ~ \frac{\prod_{z_\ell}(1 - \frac{\omega-a}{z_\ell-a})^{\nu_{z_\ell}}}{\prod_{p_\ell}(1 - \frac{\omega-a}{p_\ell-a})^{\nu_{p_\ell}}} ~ e^{i\tau(\omega-a)} \\
        &\tau = -i\left(\frac{G'(a)}{G(a)} - \sum_{z_\ell} \frac{\nu_{z_\ell}}{a-z_\ell} + \sum_{p_\ell} \frac{\nu_{p_\ell}}{a-p_\ell} \right)
    \end{aligned}
    \label{eq:Log_deriv_exp}
\end{eqnarray}
where $G(a)$ is defined as:
\begin{eqnarray}
    \begin{aligned}
        G(a) &= \frac{H(a)}{a^m} ~ &\text{if } a \neq 0, \\
        G(a)  &= \frac{\partial_\omega^m[H](0)}{m!} ~&\text{otherwise}.
    \end{aligned}
\end{eqnarray}
and $G'(a)$ is defined as:
\begin{eqnarray}
    \begin{aligned}
        G'(a) &= \frac{\partial_\omega[H](a)}{a^m}-m\frac{H(a)}{a^{m+1}}~ &\mbox{if } a\neq0, \\
        G'(a) &= \frac{\partial_\omega^{(m+1)}[H](0)}{(m+1)!} ~ &\mbox{otherwise}.
    \end{aligned}
\end{eqnarray}
We refer to Eq.~(\ref{eq:Log_deriv_exp}) as the Singularity and Zero Factorization (SZF). Let us point out the presence of the aforementioned complex exponential $e^{-i\tau\omega}$ missing from the PZF, but also of the known response $G(a)$ (the static response in the case of $a=0$). Let us also stress that in the case of an LTIS, the phase introduced by the complex exponential can be set to an arbitrary position by changing the time origin or the space-origin (for a system with coupled space and time variables). Shifting the time origin by $\tau_0$ in the temporal domain results in a multiplication by $e^{i\omega\tau_0}$ of the input $X(\omega)$ in the harmonic domain. This is tantamount to considering the transfer function $H(\omega)e^{i\omega\tau_0}$ with the same input $X(\omega)$ as depicted in Fig.~\ref{fig:fig3_to} for $H$ defined as: 
\begin{eqnarray}
    \begin{aligned}
        H(\omega) &= 5 + \frac{2 + 0.1i}{\omega - (2 + 0.3i)} \\
        &+ \frac{4 + 0.1i}{\omega - (5 + 1.4i)} + \frac{6 + 0.1i}{\omega - (6.5 + 0.7i)}
    \end{aligned}
    \label{eq:H_example_2}
\end{eqnarray}
In Fig.~\ref{fig:fig3_to} (a), the phase of the transfer function $H(\omega)$ is shown in the complex plane. In Fig.~\ref{fig:fig3_to} (b), it is plotted for $H(\omega)$ multiplied by $e^{i\omega\tau_0}$ with $\tau_0$ arbitrarily set to $\pi/6$. We show that the position of the zeros and singularities remains the same after switching from $H(\omega)$ to $H(\omega)e^{i\omega\tau_0}$, but a non-constant phase-shift is introduced in the complete complex frequency plane by the phasor $e^{i\omega\tau_0}$. It is thus possible to set $\tau_0$ in such a way that $\tau + \tau_0 = 0$, where $\tau$ is the time constant naturally appearing in the SZF in Eq.~\eqref{eq:Log_deriv_exp}.

By setting $\tau=0$, $a=0$, and appropriately defining two constants $N_0$ and $D_0$ appropriately, we recover the PZF described in the previous section by considering only $N_s$ poles and $N_z$ zeros:
\begin{eqnarray}
    \begin{aligned}
        &H(\omega) = \frac{N_0\prod_{\ell=1}^{N_z}(i\omega - iz_\ell)}{D_0\prod_{\ell=1}^{N_s}(i\omega - i p_\ell)} \\
        &N_0 = \left[\prod_{\ell=1}^{N_z}\frac{i}{z_\ell}\right] ~ \omega^m ~ G(0)\\
        &D_0 = \prod_{\ell=1}^{N_s}\frac{i}{p_\ell}
    \end{aligned}
\end{eqnarray}

\begin{figure}[h!]
    \centering
    \includegraphics[width=\columnwidth]{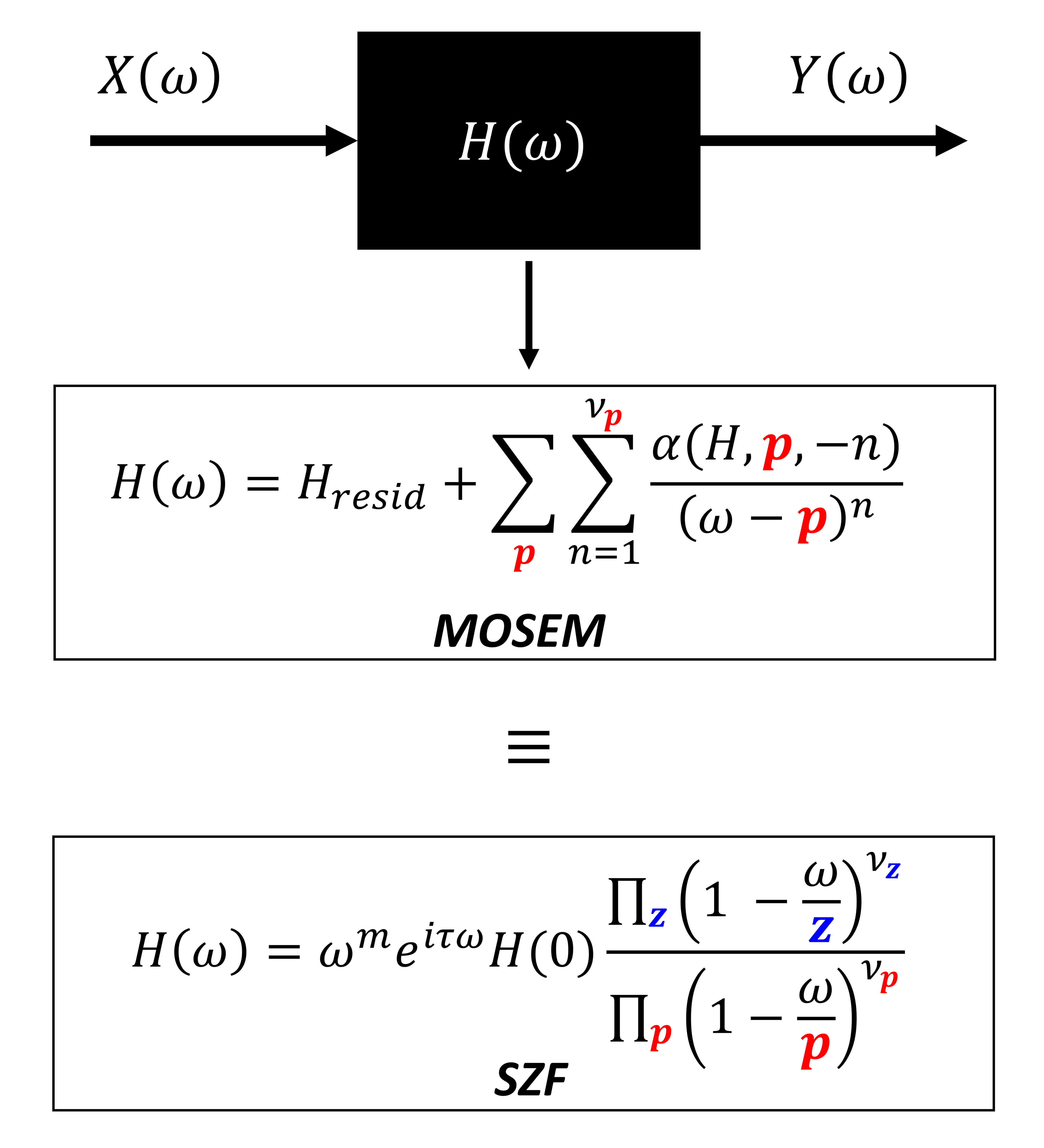}
    \caption{Any physically realistic input $X(\omega)$, output $Y(\omega)$ or transfer function $H(\omega)$ can be described using either MOSEM or the SZF. The two expressions are equivalent.}
    \label{fig:fig6_but4really}
\end{figure}
The MOSEM and SZF expressions, which are reminded in Fig.~\ref{fig:fig6_but4really}, are two equivalent means to characterize a function: (i) MOSEM expression, which relies on the behaviour in the vicinity of the singularities only and thus depend on the singularities and Laurent series coefficients associated with them, is useful to get a fast and accurate approximation of a function in the harmonic-domain; (ii) the SZF is more convenient to look at the phase and better understand the behaviour on the real frequency axis. It is easy to obtain the Laurent series coefficients, and thus MOSEM expression, from the SZF using the definition of the Laurent series coefficients:
\begin{eqnarray}
    \begin{aligned}
        &\alpha(H,p_0,-m) = \lim_{\omega \rightarrow p_0} \frac{1}{m!}\partial_\omega^{m-1} [\eta](\omega) \\
        &\eta(\omega) = (\omega-p_0)^{\nu_{p_0}}H(\omega)
    \end{aligned}
\end{eqnarray}
where $p_0$ is a pole of order $\nu_{p_0}$ of $H$, and $H(\omega)$ is given by the SZF in Eq.~(\ref{eq:Log_deriv_exp}). It is however more difficult to obtain the zeros from MOSEM expression, although they can be approximated in a specific frequency range by writing MOSEM as a rational function involving only the singularities in the selected complex frequency window and solving for the zeros of the numerator.

\section{Hermitian Symmetry and Constraints in the Harmonic Domain}

We now wish to take advantage of the physical nature of the signals to derive some constraints on the parameters of MOSEM and SZF.

\subsection{Singularity Expansion}

Let us consider a real-valued function $h(t)$ in the temporal domain, associated with a complex function $H(\omega)$. Since $h$ is real, $H$ possesses a hermitian symmetry in the complex plane~\cite{newton1966,grigoriev2013}:
\begin{eqnarray}
    H(-\omega^*)^* = H(\omega)
    \label{eq:pole_conj}
\end{eqnarray}
This property leads to constraints on the distribution of the poles and singularities in the complex frequency plane. By evaluating the complex conjugate of the singularity expansion in Eq.~(\ref{eq:f_approx}) evaluated at $-\omega^*$, it can be shown that for any pole $p$ of order $\nu_p$, $-p^*$ is also a pole of order $\nu_p$. 
\begin{figure*}
    \centering
    \includegraphics[width=0.9\textwidth]{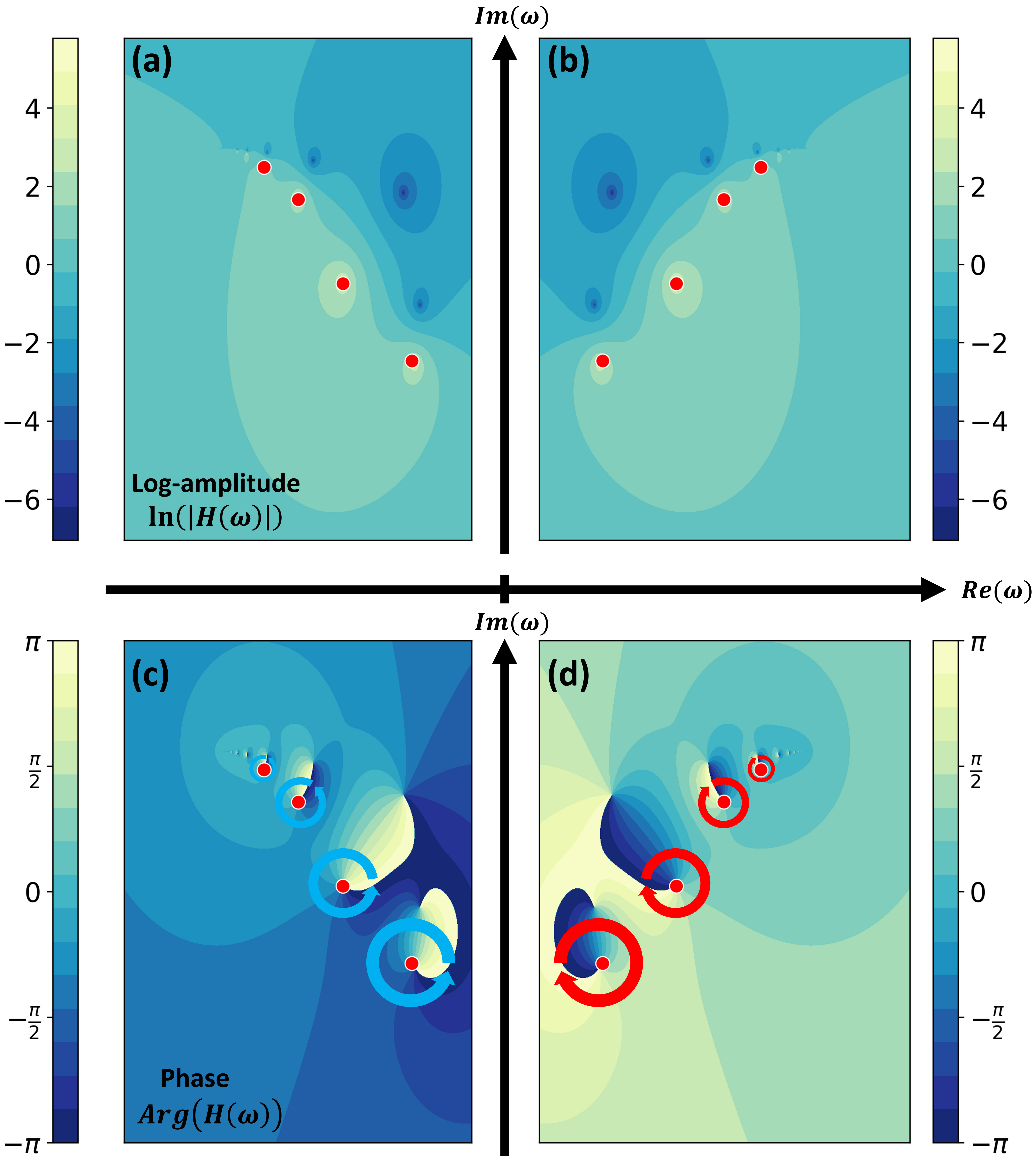}
    \caption{Complex Bode diagram of the function $H$ defined in Eq.~(\ref{eq:H_example}). (a),(b) Log-amplitude of $H$ for a complex frequency window in (a) and its symmetric window with respect to the imaginary axis in (b). (c),(d) The phase of $H$ in the same frequency window (c) and its symmetric window (d). The amplitude is symmetric with respect to the imaginary axis and results in a symmetric distribution of the poles (red points). The phase is antisymmetric, as shown with the red and blue arrows indicating a clockwise $2\pi$ phase-shift around the poles with a positive and negative real part respectively.}
    \label{fig:SPE}
\end{figure*}
We can also obtain the Laurent coefficients associated with $-p^*$ to those of $p$ \textit{via}:
\begin{eqnarray}
    \forall m\in\mathbb{Z},~\alpha(H, -p^*, -m) = -\alpha(H, p, -m)^*
    \label{eq:res_pole_conj}
\end{eqnarray}
and we show similarly that the non-resonant term $H_{\text{NR}}$ must be real. The Laurent series coefficients of $H$ at $-p^*$ are the opposite of the complex conjugate of those at $p$.
This leads to the following MOSEM expression:
\begin{eqnarray}
    \begin{aligned}
        H(\omega) &= H_{\text{NR}} + \sum_{p, ~ p\in i\mathbb{R}} \sum_{m=1}^{\nu_p} \frac{\alpha(H, p, -m)}{(w-p)^m}\\
        &+ \sum_{p, ~ Re[p]>0} \sum_{m=1}^{\nu_p} \left[\frac{\alpha(H, p, -m)}{(w-p)^m} - \frac{\alpha(H, p, -m)^*}{(w+p^*)^m} \right]
    \end{aligned}
    \label{eq:f_approx_constrained}
\end{eqnarray}
Let us point out that Eq.~(\ref{eq:res_pole_conj}) restricts $\alpha(H, p, -m)$ to $i\mathbb{R}$ if $p\in i\mathbb{R}$.
The poles and Laurent series coefficients thus always come in pairs, as shown in Fig.~\ref{fig:SPE} for the transfer function $H$ defined in Eq.~(\ref{eq:H_example}). In (a) and (b), the log-amplitude in a complex frequency window as well as its symmetric window is plotted, highlighting the symmetry of the amplitude and thus of the singularities. In (c) and (d), the same frequency windows were chosen for the phase plots. They show the antisymmetry of the phase relative to the imaginary axis in the complex frequency plane, which is tantamount to an antisymmetry of the Laurent series coefficients. Let us stress that the hermitian symmetry only arises from the fact that we have real-valued signals in the temporal domain. Considering a causal plane wave $e^{-i\omega_0 t}u(t)$, with $u$ the Heaviside step function and $\omega_0 > 0$ is not equivalent to considering that time flows backward. It only means that the variations of the phase are opposed to those of $e^{i\omega_0 t}u(t)$.

\subsection{Singularity and Zero Factorization}
If we now look at the complex conjugate of the SZF (Eq.~(\ref{eq:Log_deriv_exp})) evaluated at $-\omega^*$, we can show that if $z$ is a zero of order $\nu_z$ of $H$, then $-z^*$ is also a zero of order $\nu_z$. In addition, let us show that the Hermitian symmetry forces the time constant $\tau$ to be real-valued. If $\tau$ were complex-valued, it could be written $\tau=\tau_R + i\tau_I$, with $\tau_R$ and $\tau_I$ the real and imaginary parts of $\tau$ respectively. Since the opposite of the complex conjugate of zeros and poles are also zeros and poles of the same order, the Hermitian symmetry would thus lead to:
\begin{eqnarray}
    \begin{aligned}
        ~\omega^m~e^{i\tau_R\omega}~e^{-\tau_I\omega} = (-1)^m~\omega^m~e^{i\tau_R\omega}~e^{\tau_I\omega}
    \end{aligned}
\end{eqnarray}
This condition would not be satisfied as $\omega$ tends towards $+\infty$ unless $\tau_I=0$. Therefore, $\tau \in \mathbb{R}$. 
If we set $a=0$, the SZF thus yields the following expression of $\tau$:
\begin{eqnarray}
    \begin{aligned}
        \tau = &-i \left(\frac{G'(0)}{G(0)} - \sum_{p_\ell\in i\mathbb{R}} \frac{\nu_{p_\ell}}{p_\ell} \right) \\
        &+ 2Im\left(\sum_{z_\ell, Re[z_\ell]>0} \frac{\nu_{z_\ell}}{z_\ell} - \sum_{p_\ell, Re[p_\ell]>0} \frac{\nu_{p_\ell}}{p_\ell} \right) 
    \end{aligned}
\end{eqnarray}
with $G'(0)/G(0) \in i\mathbb{R}$. The time constant $\tau$ is thus the sum of two contributions: (1) the phase-shift introduced by the imaginary part of the singularities and the non-null zeros, (2) a constant term depending on the static response of the derivatives of $H$. In physical systems for which the space and time variables are coupled, the phase shift can be modified by moving the spatial or temporal origin. In this case, the zeros and poles are identical, and only the constant term linked to the static response is changed.

\subsection{Stability and Causality}
Stability and causality are linked but distinct principles which can both be expressed in terms of the position of the singularities in the complex plane depending on the convention used to perform a Fourier transform~\cite{Nussenzveig1972}.
Causality states that any signal $h$ must be generated at a certain time $t_h$, and that it cannot depend on its future values. If $h$ is a causal signal, it can therefore be written, using the Heaviside step function $u$, as:
\begin{eqnarray}
    \begin{aligned}
        h(t)=h(t)u(t-t_h)
    \end{aligned}
\end{eqnarray}
As long as $h$ does not diverge faster than an exponential function for a long time $t$, it can thus always be regularized using a function $h_\gamma$, $\gamma>0$, which converges:
\begin{eqnarray}
    \begin{aligned}
        &h(t) = e^{\gamma t} h(t) e^{-\gamma t} u(t) \\
        &h(t) = e^{\gamma t} h_\gamma(t)
    \end{aligned}
\end{eqnarray}
Let us point out that we set $t_h=0$ in $u$, and we can do so without losing in generality. By construction, $h_\gamma$ possesses a Fourier transform $H_\gamma$ from which the harmonic domain function $H$ associated with $h$ can be determined:
\begin{eqnarray}
    \begin{aligned}
        H(\omega) = H_\gamma(\omega - i\gamma)
    \end{aligned}
\end{eqnarray}
$h$ can be retrieved by integrating $H$ (multiplied by a complex exponential) over a horizontal line within the region of convergence of $H$, lower-bounded by the amplitude of the smallest diverging exponential function diverging faster than $h$ (see Fig.~\ref{fig:fig5_laplace_fourier}). Using the residue theorem, it can thus be shown that any pole possessing a positive imaginary part in the complex frequency plane is associated with a causal diverging exponential function in the temporal domain, \textit{i.e.} an unstable signal. Therefore, the poles of the harmonic-domain function associated with any stable signal must have a negative imaginary part. The stability of the signals can be interpreted with energy considerations. As an input signal interacts with a system, it exchanges energy with it. This leads to a modification of the input signal which result in the output signal. For a passive system, the energy is transferred from the input to the system. The output thus has a lower energy and cannot diverge if the input is stable. For the output to diverge or become unstable, it must result from a sufficient energy transfer from the system to the input, or the interaction with an already diverging input with the system. Therefore, the only way to obtain an unstable output is through a high-energy, unstable input, or an active system. In terms of singularities, this means that the singularities of the transfer function of a passive system always possess a negative imaginary part (using our Fourier transform convention). If we inject energy into the system, \textit{i.e.} the system is active, we move the singularities closer and closer to the real axis until the system is unstable and at least one singularity possess a positive imaginary part.

\section{Temporal Expressions with the Multiple-Order Singularity Expansion Method}\label{sec:temporal_exp}

\begin{figure*}
    \centering
    \includegraphics[width=\textwidth]{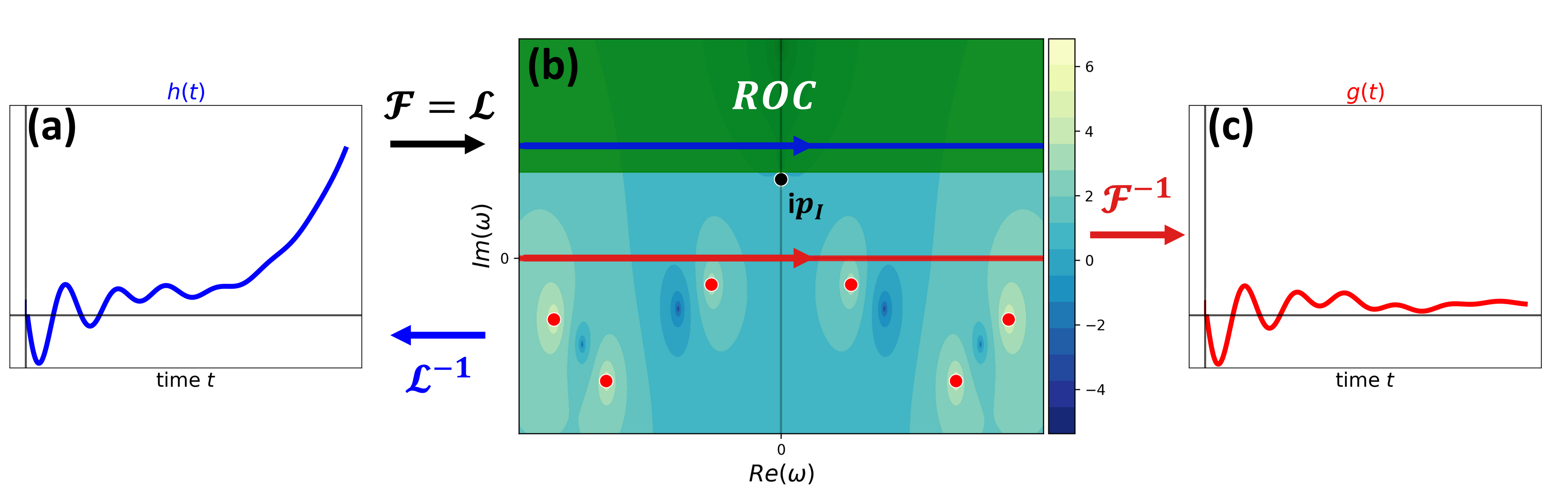}
    \caption{(a) An unstable temporal signal $h(t)$ equivalent to $e^{p_It}$ for long times, with $p_I = 0.9$. It is obtained via inverse Laplace transform of the function $H(\omega)$ from Eq.~(\ref{eq:H_example_2}) to which a pole $ip_I$ with residue $8+0.1i$ was added. (b) The harmonic-domain function $H(\omega)$, which can be obtained by Fourier or Laplace transform of $h(t)$. (c) A stable temporal function $g(t)$ obtained via inverse Fourier transform of $H$. The Laplace and Fourier transformation are equivalent since the signal is causal, \textit{i.e.} $h(t)=h(t)u(t)$. The Region of convergence (ROC) corresponds to the part of the complex frequency plane where the usual Laplace transform is defined, and within which the inverse Laplace transform should be performed to retrieve the signal $h(t)$ in (a). Below the ROC, another function would be obtained. In particular, the inverse Fourier transform is performed on the real axis, and leads to the stable function $g(t)$ in (c)}
    \label{fig:fig5_laplace_fourier}
\end{figure*}

The expansion obtained with MOSEM can be used to derive an analytical expression in the temporal domain~\cite{colom2018,bensoltane2022}. For physical systems, unstable behaviours might appear, which prevent the use of the inverse Fourier transform~\cite{valagiannopoulos2022}. It is therefore preferable to use the more general inverse Laplace transform to retrieve the temporal dynamics of a system or its response (and/or input)~\cite{Nussenzveig1972}. In this section, we derive a generalized expression of the temporal domain signal $h$ knowing its singularity expansion.
Similarly to the previous section, if $h$ is physically realistic, then causality implies that :
\begin{eqnarray}
    h(t) = h(t)u(t-t_y)
\end{eqnarray}
Let us point out that if $h$ is the response of an LTIS, the time constant $t_h$ corresponds to the time it takes for the system to interact with the input signal and produce the response.
We shift the time origin to set $t_h=0$. The Laplace transform $\mathcal{L}$ and the Fourier transform $\mathcal{F}$ are equivalent in this case:
\begin{eqnarray}
    \begin{aligned}
        \mathcal{F}[h](\omega) &= \int_{-\infty}^{+\infty} h(t)u(t)e^{i\omega t} dt \\
        &= \int_{0}^{+\infty} h(t)e^{i\omega t} dt = \mathcal{L}[h](\omega)
    \end{aligned}
\end{eqnarray}
Using these conventions, more properties regarding the inverse Laplace and Fourier transforms can be deduced. Let us consider the temporally diverging (or unstable) signal $h(t)$ increasing slower than an exponential function in Fig.~\ref{fig:fig5_laplace_fourier} (a). The Laplace or Fourier transform of $h(t)$ has at least one pole in the upper half of the complex plane. This pole, which is called $ip_I$ in Fig.~\ref{fig:fig5_laplace_fourier} (b), imposes the Region Of Convergence (ROC) of the Laplace transform of the signal (green band at the top). The inverse Laplace transform is defined as an integral over a horizontal in the complex plane. If we choose that horizontal line above the imaginary part of all the poles, within the ROC, we retrieve the original signal $h(t)$ (from (b) to (a)). Otherwise, since only the contribution of the singularities below are taken into accounts and we obtain another temporal function $g(t)$. The inverse Fourier transform is a special case of the inverse Laplace transform in which the horizontal line is the real frequency axis. Therefore, performing an inverse Fourier transform on the harmonic-domain function of an unstable signal does not allow the retrieval of the temporal signal (from (b) to (c) in Fig.~\ref{fig:fig5_laplace_fourier}). It is therefore necessary to perform, in general, an inverse Laplace transform over a horizontal line above the real axis and any potentially unstable pole.

Let us now calculate the temporal-domain function $h(t)$ associated with the harmonic-domain function $H(\omega)$. We apply MOSEM to $H$ (see Eq.~\ref{eq:f_approx}) and derive the inverse Laplace transform of every term:
\begin{eqnarray}
    \label{eq:transfer_temporal}
    \begin{aligned}
        h(t) &= H_{\text{NR}}\delta(t) \\
        &+ \left(\sum_{p} \sum_{m=1}^{\nu_p} \frac{(-i)^n\alpha(H, p, -m)}{(n-1)!}t^{n-1}e^{-ipt} \right) u(t)
    \end{aligned}
\end{eqnarray}

\subsection{Response to a Sinusoidal Input}
In many problems, the main focus is the global temporal response of the LTIS at a specific frequency. In these situations, the characterization of the system in the harmonic domain is sufficient since it is almost equivalent to a study of the temporal permanent-regime. We expect the response of a stable system to an excitation to be composed of that same input, scaled and phase-shifted in the permanent-regime, as well as exponentially decaying functions in the transient-regime ~\cite{bensoltane2022}. It is worth demonstrating this known result by deriving the response $y(t)$ to a sinusoidal input $x(t) = Re[e^{-i\omega_0t}u(t - t_e)]$ of an LTIS of impulse response $h(t)$. Since $y$ and $x$ are real, $h$ must be real, which leads to:
\begin{eqnarray}
    \begin{aligned}
        y(t) &= (h \ast Re\left[x\right])(t) = Re\left[(h \ast x)(t)\right] \\
        &= Re\left[\mathcal{L}^{-1}[H.X]\right] = Re\left[\mathcal{L}^{-1}[Y]\right] \\
    \end{aligned}
\end{eqnarray}
The response to a sinusoidal input is the real part of the response to a causal plane wave $t \mapsto e^{-i\omega_0t}u(t)$. In the harmonic domain, $Y$ and $X$ can be written as:
\begin{eqnarray}
    \begin{aligned}
        &Y(\omega) = X(\omega)H(\omega) \\
        &X(\omega) = \frac{i}{\omega - \omega_0}
    \end{aligned}
\end{eqnarray}
Therefore, the set of isolated singularities of the harmonic response includes the singularities of the transfer function with the same order, as well as $\omega_0$ of order 1 introduced by the harmonic input signal $X$. Applying MOSEM to $Y$ gives:
\begin{eqnarray}
    Y(\omega) = Y_{\text{NR}} + Y_{\text{R}}(\omega)
    \label{eq:Y_harmonic0}
\end{eqnarray}
with
\begin{eqnarray}
    \begin{aligned}
        &Y_{\text{NR}} = Y(0) - \sum_{p} \sum_{m=1}^{\nu_p} \frac{\alpha(Y, p, -m)}{(-p)^m} + \frac{\alpha(Y,\omega_0,-1)}{\omega_0}, \\
        &Y_{\text{R}}(\omega) = \sum_{p} \sum_{m=1}^{\nu_p} \frac{\alpha(Y, p, -m)}{(w-p)^m} + \frac{\alpha(Y,\omega_0,-1)}{\omega - \omega_0}
    \end{aligned}
    \label{eq:Y_harmonic}
\end{eqnarray}
The calculation of the Laurent series coefficients of $Y$ leads to:
\begin{eqnarray}
    \begin{aligned}
        &\alpha(Y, p, n) =  -\sum_{\ell=-\nu_p}^{n} \frac{i\alpha(H,p,\ell)}{(\omega_0 - p)^{n-\ell+1}}.
    \end{aligned}
\end{eqnarray}
for the singularity $p$ of order $\nu_p$. In addition, the residue of $Y$ at $\omega_0$ is proportional to the transfer function evaluated at the plane wave frequency $\omega_0$:
\begin{eqnarray}
    \begin{aligned}
        &\alpha(Y, \omega_0, -1) = \lim_{\omega \rightarrow \omega_0} (\omega - \omega_0)X(\omega)H(\omega), \\
        &\alpha(Y, \omega_0, -1) = iH(\omega_0)
    \end{aligned}
\end{eqnarray}
By replacing the corresponding terms in Eqs.~(\ref{eq:Y_harmonic0},\ref{eq:Y_harmonic}), we obtain:
\begin{eqnarray}
    \begin{aligned}
        Y(\omega) &= Y_{\text{NR}} + i\frac{H(\omega_0)}{\omega - \omega_0} + \sum_{p} \sum_{m=1}^{\nu_p} \frac{\alpha(Y, p, -m)}{(w-p)^m}
    \end{aligned}
\end{eqnarray}
It follows that the inverse Laplace transform of $Y$ can be written as the sum of a Dirac's delta function multiplied by $Y_{\text{NR}}$, an expansion on the dynamic states similarly to the resonant term in Eq.~(\ref{eq:transfer_temporal}) after replacing $\alpha(H, p, n)$ by $\alpha(Y, p, n)$, and the scaled and phase-shifted sinusoidal function $t \mapsto e^{-i\omega_0t}$:
\begin{eqnarray}
    \begin{aligned}
        \mathcal{L}^{-1}[Y](t) &= Y_{\text{NR}}\delta(t) + H(\omega_0)e^{-i\omega_0t} u(t) \\
        &+\sum_{p} \sum_{m=1}^{\nu_p} \alpha(Y, p, -m) f_{m, p}(t)
    \end{aligned}
\end{eqnarray}
with
\begin{eqnarray}
    f_{m,p}(t) = \frac{(-i)^m}{(m-1)!}t^{m-1}e^{-ipt}~u(t)
\end{eqnarray}
The initial value theorem imposes that $Y_{\text{NR}}$ must be null if $x$ is a continuous input. Since $\lim_{t \rightarrow 0^+} y(t) = 0$, \textit{i.e.} the system cannot immediately respond to a physical signal, the initial value theorem applied to $y$ gives:
\begin{eqnarray}
    \begin{aligned}
        &0 = \lim_{t \rightarrow 0^+} y(t) = \lim_{\omega \rightarrow +\infty} i\omega Y(\omega)
    \end{aligned}
\end{eqnarray}
It follows that the constant, non-resonant term must be equal to 0, and that the sum of the residues of $Y$ is null:
\begin{eqnarray}
    \begin{aligned}
        &\left( \omega Y_{\text{NR}} \rightarrow 0 \right) \Rightarrow (Y_{\text{NR}} = 0) \\
        &0 = \alpha(Y,\omega_0,-1) + \sum_p \alpha(Y, p, -1)
    \end{aligned}
\end{eqnarray}

We obtain the aforementioned expected result: the response $y$ of a stable system to a causal sinusoidal input is only composed of the scaled and phase-shifted input in the permanent regime. In the transient regime, $y$ must be expanded on the set of dynamic states $f_{p,n}$ of the system:
\begin{eqnarray}
    \begin{aligned}
        \mathcal{L}^{-1}[Y](t) &= H(\omega_0)e^{-i\omega_0t} u(t) \\
        &+ \sum_{p} \sum_{m=1}^{\nu_p} \alpha(Y, p, -m) f_{m, p}(t)
    \end{aligned}
\end{eqnarray}
This expansion remains valid in the case of an unstable system, although the concept of transient and permanent regime would no longer hold. In this case, the dynamic states would hold information regarding the diverging speed of $y$.

\section{Conclusion}
We extended in this work the singularity expansion method to the general case of multiple order singularities in the complex frequency plane. Starting from simple considerations regarding the physical nature of the signals, we detailed the derivation of  a more general singularity expansion, from which we deduced the singularity and zero factorization of a function. We calculated the exact temporal-domain expression of the response or the impulse response using the inverse Laplace transform of the generalized singularity expansion. By considering the case of the response to a sinusoidal input, we show that the singularities form a natural basis for the expansion of temporal responses in the transient regime, while they only influence the amplitude of the sinusoidal signal in the permanent regime. Finally, we inferred the constraints put on the poles and zeros in the complex plane for physically realistic signals possessing a Hermitian symmetry in the harmonic domain. Furthermore, causality was assumed to discriminate between stable and unstable poles based on the sign of their imaginary part. We believe that the multiple-order singularity expansion method will find applications as a means to unveil specific properties of linear systems and their response by linking physical phenomenon to the distribution of the singularities and zeros, but also as a numerical tool to obtain highly accurate approximations of functions knowing only some of their singularities.  Further works will explore the richness of these properties, in both the harmonic and time domains.   

\begin{acknowledgments}
This work was funded by the French National Research Agency ANR
Project DILEMMA (ANR-20-CE09-0027). The authors thank Eve-line Bancel for the fruitful discussions.
\end{acknowledgments}

\bibliography{apssamp}
\bibliographystyle{iopart-num.bst}

\end{document}


\preprint{APS/123-QED}

\title{Supporting Information on ``Multiple-Order Singularity Expansion Method''}

\author{Isam Ben Soltane}
 \email{isam.ben-soltane@fresnel.fr}
\affiliation{Aix Marseille Univ, CNRS, Centrale Marseille, Institut Fresnel, 13013 Marseille, France}
\author{Rémi Colom}
\affiliation{CNRS, CRHEA, Université Côte d’Azur, 06560 Valbonne, France}
\author{Félice Dierick}
\affiliation{Aix Marseille Univ, CNRS, Centrale Marseille, Institut Fresnel, 13013 Marseille, France}
\author{Brian Stout}%
\affiliation{Aix Marseille Univ, CNRS, Centrale Marseille, Institut Fresnel, 13013 Marseille, France}
\author{Nicolas Bonod}%
\email{nicolas.bonod@fresnel.fr}
\affiliation{Aix Marseille Univ, CNRS, Centrale Marseille, Institut Fresnel, 13013 Marseille, France}

\date{\today}

\maketitle

\section{Refractive Index of Silver}

The reflection coefficient of a thin layer of silver illuminated from one-side at normal incidence is used as an example in the main manuscript. It has the following expression:
\begin{eqnarray}
    \begin{aligned}
        H(\omega) = r(\omega) - \frac{t(\omega)t'(\omega)r(\omega)e^{2i\omega \frac{n(\omega)d}{c}}}{1 - r(\omega)^2e^{2i\omega \frac{n(\omega)d}{c}}}
    \end{aligned}
    \label{eq:H_example_SI}
\end{eqnarray}
where $d=70$~nm is the thickness of the silver layer, $c$ is the speed of light in the air, $n(\omega)$ is the refractive index of silver, $r(\omega)=(n(\omega)-1)/(n(\omega)-1)$ is the Fresnel reflection coefficient at the air/silver interface, and $t(\omega)=2/(n(\omega)+1)$ and $t'(\omega)=2n(\omega)/(n(\omega)+1)$ The expression of $n(\omega)$ is obtained by analytical continuation into the complex frequency plane of the permittivity using a Drude-Lorentz which fits experimental data~\cite{johnson1972} on the real frequency axis. The expression of the relative electric permittivity $\epsilon(\omega)$ is the following:
\begin{eqnarray}
    \begin{aligned}
        \epsilon(\omega) = \epsilon_{\infty} + &i\frac{G_0}{\omega} - \sum_{\ell=1}^{3} \frac{\omega_{i,\ell}^2}{\omega^2 + i\Gamma_{i,\ell}\omega} \\
        &- \sum_{m=1}^{7} \frac{is_{1,m}\Gamma_m\omega + s_{2,m}\omega_m^2}{(\omega^2 - \omega_m^2) + i\Gamma_m\omega}
    \end{aligned}
\end{eqnarray}
with $\epsilon_{\infty}=0.763$, $G_0=0.704$, $\omega_{i,1}=1.870$, $\omega_{i,2}=2.295$, $\omega_{i,3}=13.54$, $\Gamma_{i,1}=0.400$, $\Gamma_{i,2}=0.800$, $\Gamma_{i,3}=1.000 \times 10^{-4}$, $s_{1,1}=0.679$, $s_{1,2}=-0.352$, $s_{1,3}=-1.534$, $s_{1,4}=-1.154$, $s_{1,5}=-0.136$, $s_{1,6}=0.837$, $s_{1,7}=1.065$, $\Gamma_1=0.346$, $\Gamma_2=0.753$, $\Gamma_3=0.698$, $\Gamma_4=2.196$, $\Gamma_5=1.909$, $\Gamma_6=1.406$, $\Gamma_7=0.020$, $s_{2,1}=1.493$, $s_{2,2}=0.021$, $s_{2,3}=0.016$, $s_{2,4}=0.813$, $s_{2,5}=0.260$, $s_{2,6}=0.163$, $s_{2,7}=0.216$, $\omega_1=0.852$, $\omega_2=2.747$, $\omega_3=6.160$, $\omega_4=6.856$, $\omega_5=8.512$, $\omega_6=9.943$ and $\omega_7=10.786$. Let us point out that the low values of the parameters are due to the multiplication of the frequency by $10^{-15}$ in the calculations.
The refractive index used in Eq.~(\ref{eq:H_example_SI}) is defined as the square root of the relative permittivity:
\begin{eqnarray}
    \begin{aligned}
        n(\omega) = \sqrt{\epsilon(\omega)}
    \end{aligned}
\end{eqnarray}

\section{Multiple-Order Singularity Expansion Method}

\subsection{Deriving the Expansion}

We wish to derive a global expansion of a complex function in terms of its singularities. The emphasis is put on the harmonic-domain function $H$ associated with the temporal signal $h$. In what follows, $H$ is assumed to be meromorphic, \textit{i.e.} holomorphic everywhere on $\mathbb{C}$ except for a set of points $\mathcal{P}$ which is the set of its isolated singularities. 
Let us consider a domain of $\mathcal{C}$ containing $\omega$, but no singularity of $H$, bounded by a curve $\gamma_C$. By definition, $H$ is holomorphic in this domain, and the Cauchy Integral Theorem gives:
\begin{eqnarray}
    \label{eq:Cauchy}
    \int_{\gamma_C} \frac{H(z)}{z-\omega} dz = 2i\pi H(\omega)
\end{eqnarray}
It is therefore possible to obtain $H$ at the frequency $\omega$ from the knowledge of $H$ around this frequency, in a small domain. As the size of the domain increases, more and more singularities are contained within it, and we have to make use of the residue theorem to fully characterize $H$ at $\omega$.

\begin{figure}[ht!]
    \centering
    \includegraphics[width=0.95\columnwidth]{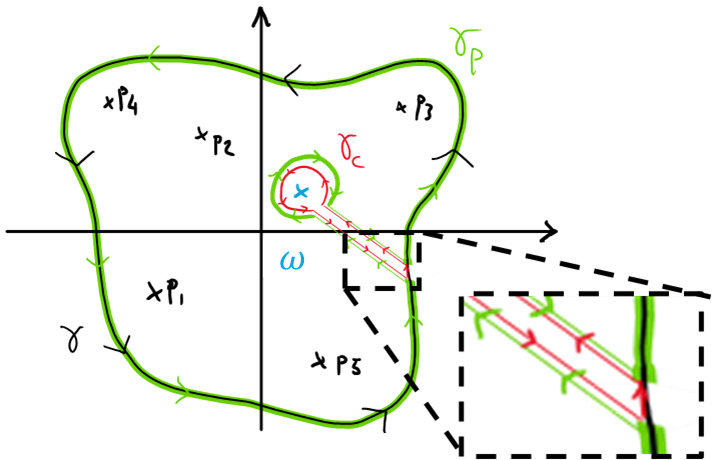}
    \caption{Separation of the integration curve $\gamma$ (black curve) into a circle $\gamma_C$ (red curve) around $\omega$ (blue cross) and a curve $\gamma_P$ (green curve) around the singularities. The zoom in is used to stress the complementary of the red and green curves to produce the black one.}
    \label{fig:sep_gamma}
\end{figure}

Let us now consider another closed curve $\gamma$ (black curve in figure~\ref{fig:sep_gamma}) around a domain $\mathcal{D}$ containing a subset of the singularities of $H$, and $\omega \in \mathcal{D}$. The integral over $\gamma$ can be expressed as the integral over $\gamma_P$, the closed curve surrounding the singularities and excluding $\omega$ (red curve in \ref{fig:sep_gamma}), plus the integral over $\gamma_C$, a closed curve around $\omega$ containing no singularities (green curve in figure~\ref{fig:sep_gamma}).

Therefore, the generalized version of the Cauchy's integral formula~\cite{needham_visual_1997} leads to:
\begin{eqnarray}
    \int_{\gamma} \frac{H(z)}{z-\omega} dz = \int_{\gamma_C} \frac{H(z)}{z-\omega} dz + \int_{\gamma_P} \frac{H(z)}{z-\omega} dz
\end{eqnarray}

The integral over $\gamma_C$ is $2i\pi H(\omega)$ as stated in Eq.~\ref{eq:Cauchy}. The integral over $\gamma_P$ can be calculated using the residue theorem for the function $R$ defined as: 
\begin{eqnarray}
    \begin{aligned}
        R: z\in\mathbb{C} \mapsto \frac{H(z)}{z-\omega}
    \end{aligned}
\end{eqnarray}
R is holomorphic everywhere on $\mathbb{C}$ except on its isolated singularities:
\begin{eqnarray}
    \int_{\gamma_P} \frac{H(z)}{z-\omega} dz = 2i\pi \sum_{p} Res(R,p),
\end{eqnarray}
where the singularities $p$ considered are those within the integration curve $\gamma$
The integration curve along $\gamma_P$ is shown in green in figure~\ref{fig:sep_gamma}. 
The following expression is obtained:
\begin{eqnarray}
    \begin{aligned}
        \frac{1}{2i\pi}\int_{\gamma} \frac{H(z)}{z-\omega} dz &= H(\omega) + \sum_{p} Res(R,p)
    \end{aligned}
\end{eqnarray}

The residue of $R$ for the singularity $p$ can be expressed as a combination of the coefficients of negative order in the Laurent expansion of the function $H$ around $p$, and the coefficients of the Taylor or Laurent series expansion of $D: z \mapsto \frac{1}{z-w}$. The Taylor series coefficients of $D$ can be explicitly calculated by definition, leading to an expression of $Res(R, p)$ where only the Laurent series coefficients of $H$ are unknown.
The Laurent series expansion of the transfer function $H$ for $z$ in a small punctured disk around $p$ is written as~\cite{yger2001}:
\begin{eqnarray}
    \label{eq:H_laurent}
    H(z) = \sum_{\ell=-\nu_p}^{\infty} \alpha(H, p, \ell) . (z - p)^\ell
\end{eqnarray}
where $\nu_p$ is the order of the singularity $p$ and $\alpha(H,p, \ell)$ is the Laurent series coefficient of $H$ of order $\ell$ for the singularity $p$.
Similarly, the expression of the Taylor expansion of $D: z \mapsto \frac{1}{z - \omega}$ is:
\begin{eqnarray}
    \label{eq:D_laurent}
    \begin{aligned}
        D(z) &= \sum_{k=0}^{\infty} \beta(D, p, k) . (z - p)^k, \\
        &\beta(D, p, k) = \frac{\partial_{z}^k(D)(p)}{k!} = -\frac{1}{(\omega-p)^{k+1}},
    \end{aligned}
\end{eqnarray}
 $\beta(D,p,m)$ being the Laurent series coefficient of $D$ of order $m$ around the complex frequency $p$ (which is not a pole of $D$).
$R$ is meromorphic and its singularities, other than $\omega$, are the same as $H$ with the same order. Therefore, its Laurent series expansion near the singularity $p$ is:
\begin{eqnarray}
    \label{eq:F_Laurent}
    R(z) = \sum_{n=-\nu_p}^{\infty} \alpha(R, p, n) . (z - p)^n
\end{eqnarray}
Since $R(z)=H(z)D(z)$, Eqs.~(\ref{eq:H_laurent},~\ref{eq:D_laurent}) can also be used to provide an alternative expression:
\begin{eqnarray}
    \begin{aligned}
        R(z) &= \sum_{\ell=-\nu_p}^{n}\sum_{k=0}^{\infty}\alpha(H, p, \ell) . \beta(D, p, k)~(z-p)^{\ell+k}
    \end{aligned}
    \label{eq:F_Laurent_prod}
\end{eqnarray}
Using:
\begin{eqnarray}
    \forall \ell \in \mathcal{Z}, && (z-p)^n = (z-p)^\ell(z-p)^{n-\ell}
\end{eqnarray}
as well as the unicity of the Laurent series expansion, the terms associated with exponent $n = \ell+m$ in Eq.~\ref{eq:F_Laurent_prod} must be linked to the Laurent series coefficient of order $n$ in Eq.~\ref{eq:F_Laurent}:
\begin{eqnarray}
    \alpha(R, p, n) = \sum_{\ell=-\infty}^{+\infty} \alpha(H, p, \ell) . \beta(D, p, k=n-\ell)
\end{eqnarray}

Since $D$ is holomorphic $m\geq0 \Rightarrow \ell\leq n$. Similarly, the order of $p$ for $H$ imposes that $\ell\geq-\nu_p$:
\begin{eqnarray}
    \label{eq:Laurent_Prod}
    \alpha(R, p, n) = \sum_{\ell=-\nu_p}^{+n} \alpha(H, p, \ell) . \beta(D, p, n-\ell)
\end{eqnarray}

The Taylor series coefficient $\beta(D, p, m)$ was provided in Eq.~(\ref{eq:D_laurent}):
\begin{eqnarray}
    \beta(D, p, k) = \frac{\partial_{z}^k(D)(p)}{k!} = -\frac{1}{(\omega-p)^{k+1}}
\end{eqnarray}
which leads to:
\begin{eqnarray}
    \alpha(R, p, n) = - \sum_{\ell=-\nu_p}^{n} \frac{\alpha(H, p, \ell)}{(\omega-p)^{n-\ell+1}}
\end{eqnarray}
In particular, since $Res(R, p) = \alpha(R, p, -1)$,
\begin{eqnarray}
    Res(R, p) =  - \sum_{\ell=-\nu_p}^{-1} \frac{\alpha(H, p, \ell)}{(\omega-p)^{-\ell}}
\end{eqnarray}
For convenience, the residue is written with $m=-\ell$:
\begin{eqnarray}
    Res(R, p) =  - \sum_{m=1}^{\nu_p} \frac{\alpha(H, p, -k)}{(\omega-p)^{k}}
\end{eqnarray}
Finally, the following expression of $H$ is obtained:
\begin{eqnarray} 
    \label{eq:f_ex}
    H(\omega) = \frac{1}{2i\pi}\int_{\gamma} \frac{H(z)}{z-\omega} dz + \sum_{p} \sum_{m=1}^{\nu_p} \frac{\alpha(H, p, -m)}{(\omega-p)^m}
\end{eqnarray}

If $H$ is asymptotically well-behaved, \textit{i.e.} it does not diverge faster than $|\omega|$ when $\omega$ tends towards $\infty$ in the complex plane, then it can be written as the sum of a constant, a non-resonant term and a frequency-dependant resonant term. Under the assumptions that $|\frac{H(z)}{z}| \xrightarrow[|z|\rightarrow \infty]{} 0$, one obtains:
\begin{eqnarray}
    \begin{aligned}
        &\left| \int_{\mathcal{C}(\rho)} \frac{H(z)}{z-\omega} dz - \int_{\mathcal{C}(\rho)} \frac{H(z)}{z-a} dz \right| \\
        &\leq \int_{\mathcal{C}(\rho)} \left| H(z)\left[\frac{1}{z-\omega} - \frac{1}{z-a}\right] \right| dz \\
        &\leq 2\pi \frac{\max_{\theta \in [0, 2\pi]} \left|H(\rho e^{i\theta})\right| |\omega|\rho}{(\rho - |a|)(\rho - |\omega|)}  \xrightarrow[\rho \rightarrow \infty]{} 0 \\
    \end{aligned}
    \label{eq:abs_diff_ASE}
\end{eqnarray}
where $\mathcal{C}(\rho)$ is the circle of radius $\rho$ centered at $0$. In general, $a=0$ if $0$ is not a singularity of $H$.
Thus,
\begin{eqnarray}
    \int_{\mathcal{C}(\rho)} \frac{H(z)}{z-\omega} dz \xrightarrow[\rho \rightarrow \infty]{} \int_{\mathcal{C}(\rho)} \frac{H(z)}{z-a} dz
\end{eqnarray}
In addition, setting $\omega=a$ in Eq.~(\ref{eq:f_ex}) leads to:
\begin{eqnarray}
    \begin{aligned}
        \frac{1}{2i\pi}\int_{\mathcal{C}(\rho)} \frac{H(z)}{z-a} dz &= H(a) \\
        &- \sum_{p} \sum_{m=1}^{\nu_p} \frac{\alpha(H, p, -m)}{(a-p)^m}
    \end{aligned}
\end{eqnarray}
Therefore, when $\rho$ is large enough, \textit{i.e.} when the closed curve $C(\rho)$ around the singularities and $\omega$ tends towards $+\infty$:
\begin{eqnarray}
    \label{eq:f_approx}
    H(\omega) \approx H_{\text{NR}} + H_{\text{R}}(\omega),
\end{eqnarray}
where
\begin{eqnarray}
    \label{eq:f_approx_NR}
    H_{\text{NR}} = H(a) - \sum_{p} \sum_{m=1}^{\nu_p} \frac{\alpha(H, p, -m)}{(a-p)^m}
\end{eqnarray}
and
\begin{eqnarray}
    \label{eq:f_approx_R}
    H_{\text{R}}(\omega) = \sum_{p} \sum_{m=1}^{\nu_p} \frac{\alpha(H, p, -m)}{(w-p)^m}.
\end{eqnarray}
For $\rho = \infty$, all the singularities are contained within the interior of the integral curve. Eq.~(\ref{eq:f_approx}) becomes an equality:
\begin{eqnarray}
    \begin{aligned}
        H(\omega) &= H(a) - \sum_{p} \sum_{m=1}^{\nu_p} \frac{\alpha(H, p, -m)}{(a-p)^m} \\
        &+ \sum_{p} \sum_{m=1}^{\nu_p} \frac{\alpha(H, p, -m)}{(w-p)^m}
    \end{aligned}
    \label{eq:ASE_exp}
\end{eqnarray}
Let us point out that for most physically realistic signals $h$, the associated transfer function $H$ is indeed asymptotically well-behaved. Thus, the previous approximation is usually an exact expansion of $H$. The Multiple-Order Singularity Expansion Method (MOSEM) refers to the use of Eq.~(\ref{eq:ASE_exp}) to expand a function $H$. If $|\frac{H(z)}{z}|$ does not converge, the error introduced by MOSEM can be estimated by considering the absolute difference between $H_{\text{NR}}$ and the integral term from Eq.~(\ref{eq:f_ex}):
\begin{eqnarray}
    \begin{aligned}
        &\left|{\int_{\mathcal{C}(\rho)} (\frac{H(z)}{z-\omega} - \frac{H(z)}{z})} dz\right| \\
        &\leq 2\pi \frac{\max_{\theta \in [0, 2\pi]} \left|H(\rho e^{i\theta})\right| |\omega|\rho}{(\rho - |a|)(\rho - |\omega|)}\\
    \end{aligned}
    \label{eq:approx_error_formula}
\end{eqnarray}

\subsection{Constraints on the Poles}

We can use the singularity expansion of $H$ to deduce constraints on its singularities. If $h(t)$ is a real signal, than $H$ must have a Hermitian symmetry:
\begin{eqnarray}
    \begin{aligned}
        H(\omega) = H(-\omega^*)^*
    \end{aligned}
\end{eqnarray}
which leads to the following conditions if we apply MOSEM to $H$:
\begin{eqnarray}
    \begin{aligned}
        H_{\text{NR}} &= H_{\text{NR}}^* \\
        \sum_{p} \sum_{m=1}^{\nu_p} \frac{\alpha(H, p, -m)}{(w-p)^m} &= \sum_{p} \sum_{m=1}^{\nu_p} (-1)^m\frac{\alpha(H, p, -m)^*}{(w+p^*)^m}
    \end{aligned}
    \label{eq:hermitian_symm_poles}
\end{eqnarray}
The non-resonant term must thus be real, and we can show that the second equality implies that the opposite of the complex conjugate of the poles are also poles. Since $K:\omega \mapsto H(-\omega^*)^*$ is meromorphic and $K=H$, we can immediately infer that the poles of $K$ and $H$ are identical. In addition, Eq.~(\ref{eq:hermitian_symm_poles}) shows that for every pole $p$ of $H$, $-p^*$ is a pole of $K=H$. Finally, a Laurent series expansion of $H$ and $K$ around $-p^*$ leads to the following relation for the Laurent series coefficients of $H$ at $p$ and $-p^*$:
\begin{eqnarray}
    \begin{aligned}
        &\forall p\in\mathcal{P(H)},~\forall{m\in\mathbb{Z}}, \\
        &~\alpha(H, -p^*, -m) = -\alpha(H, p, -m)^*
    \end{aligned}
\end{eqnarray}

\subsection{Accuracy of MOSEM}

\begin{figure}
    \centering
    \includegraphics[width=0.95\columnwidth]{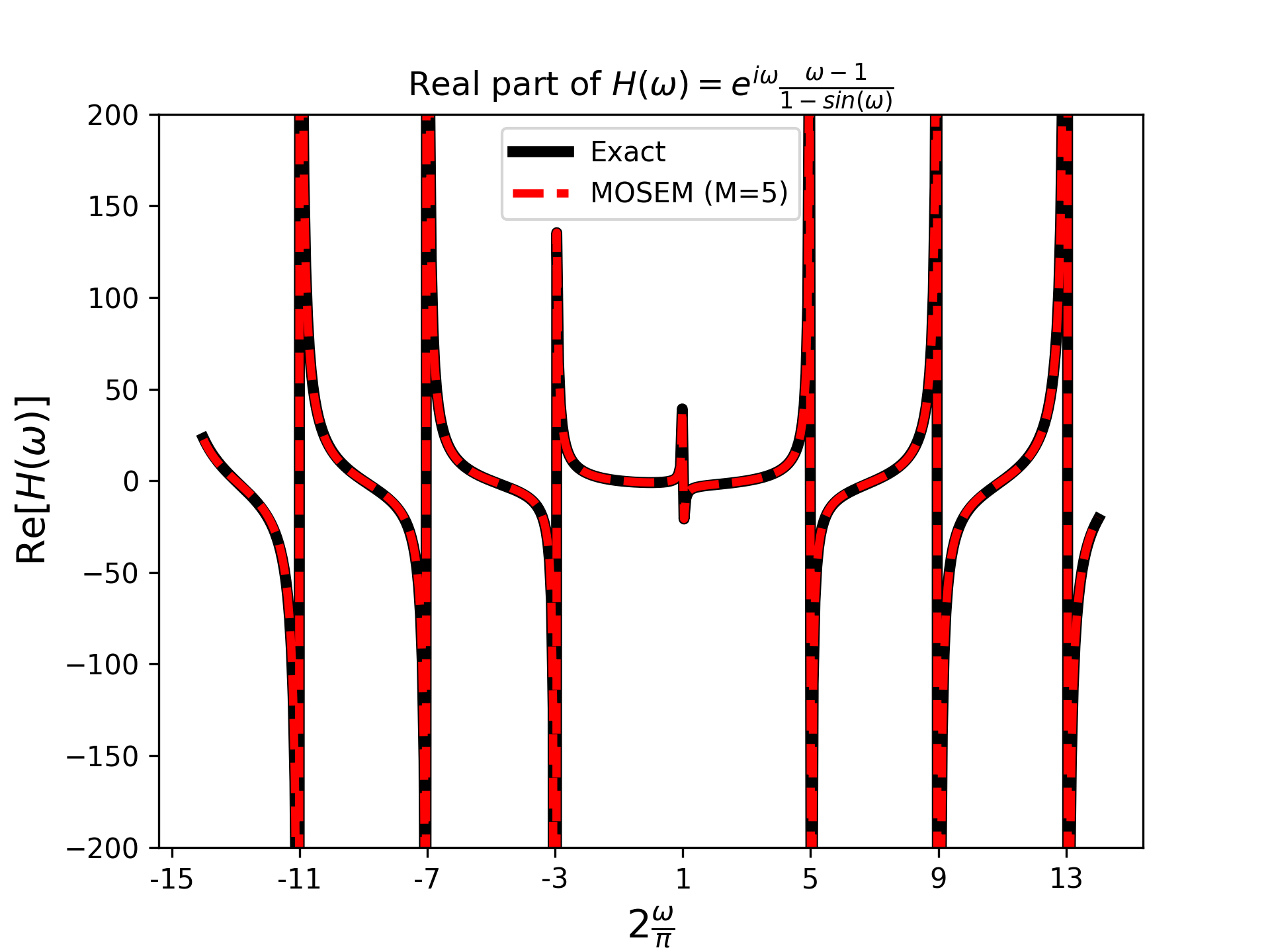}
    \caption{Real part of the function $H$ defined in Eq.~(\ref{eq:test_function_fact}). The poles of $H$ are the complex frequencies $\omega_p^{(n)}=2n\pi + \pi/2$. The exact expression in black is compared to the truncated MOSEM in red: $-M \leq n \leq M$, with $M=5$.}
    \label{fig:order2_harmonic}
\end{figure}

The transfer function of an LTIS usually has poles of order 1 only. However, the input and output signals can be shaped freely and could possess poles of arbitrary order, in which case MOSEM would prove useful. Let us consider the case of a meromorphic function with poles of order 2:
\begin{eqnarray}
    H(\omega) = e^{\text{i}\omega}\frac{\omega - 1}{1 - \sin(\omega)}
    \label{eq:test_function_fact}
\end{eqnarray}
By studying the functions $G_n$ defined as $G_n(\omega) = (\omega - \omega_p^{(n)})^2 H(\omega)$ with $\omega_p^{(n)} = (2n + 1)\pi/2$ a pole of $H$, as well as the first derivative of $G$, we can calculate the second and first negative order Laurent series coefficients of $H$:
\begin{eqnarray}
    \begin{aligned}
        &\alpha(H, \omega_p^{(n)}, -2) = 2e^{\text{i}\omega}(\omega_p^{(n)} - 1) \\
        &\alpha(H, \omega_p^{(n)}, -1) = Res(H, \omega_p^{(n)}) = 2e^{\text{i}\omega}
    \end{aligned}
\end{eqnarray}
Using these expressions, we apply MOSEM to $H$ wih $a=0$, and truncate the infinite sum over $n$ using the truncation number $M$
\begin{eqnarray}
    \begin{aligned}
        H(\omega) &\approx H(0) \\
        &+ \sum_{n=-M}^{M} 2e^{\text{i}\omega_p^{(n)}} \left[ \frac{1}{\omega_p^{(n)}} - \frac{\omega_p^{(n)} - 1}{(\omega_p^{(n)})^2}\right] \\
        &+ ~ \sum_{n=-M}^{M} 2e^{\text{i}\omega_p^{(n)}} \left[ \frac{1}{\omega - \omega_p^{(n)}} + \frac{\omega_p^{(n)} - 1}{(\omega - \omega_p^{(n)})^2}\right]
    \end{aligned}
\end{eqnarray}
We show the accuracy of MOSEM with $M=5$ in Fig.~\ref{fig:order2_harmonic}, where we compare it to the expression of $H$ from Eq.~(\ref{eq:test_function_fact}).

\section{Singularity and Zero Factorization in the Harmonic domain}

\subsection{General Expression with the Singularity Expansion}

We now use MOSEM to derive a factorized form of $H$ using its zeros $z_m$ of order $\nu_{z_m}$, and its singularities $p_\ell$ of order $\nu_{p_\ell}$. If $0$ is a zero of order $m$ of $H$, \textit{i.e.} $\forall m \leq m-1,~\partial_\omega^n[H](0)=0$, then the Hadamard factorization theorem tells us that $H$ can be expressed in terms of another meromorphic function $G$ which does not have $0$ as a zero:
\begin{eqnarray}
    \label{eq:weierstrass_fact}
    \begin{aligned}
        H(\omega) = \omega^m~G(\omega)
    \end{aligned}
\end{eqnarray}
The singularities and zeros (other than $0$) of $H$ and $G$, as well as their orders, are identical. Applying MOSEM (Eq.~(\ref{eq:ASE_exp})) to $G$ yields:
\begin{eqnarray}
    \begin{aligned}
        G(\omega) &= G(a) - \sum_{p_\ell} \sum_{m=1}^{\nu_{p_\ell}} \frac{\alpha(G, p_\ell, -m)}{(a-p_\ell)^m} \\
        &+ \sum_{p_\ell} \sum_{m=1}^{\nu_{p_\ell}} \frac{\alpha(G, p_\ell, -m)}{(\omega-p_\ell)^m}.
    \end{aligned}
\end{eqnarray}
The complex log-derivative of $G$ is the meromorphic function $F$ defined as $F = \frac{G'}{G}$, with $G'=\partial_z(G)$ the first order derivative of $G$ relative to the complex variable $\omega$, which is also meromorphic.
Using the Laurent and Taylor expansions of $G$ and $G'$, it can be shown that the singularities and of $F$ are the zeros and singularities of $H$. Their order is $1$, and the associated residues are the orders of the zeros and singularities of $G$. Let us first consider a singularity $p$ of $G$, of order $\nu_p$, and $\omega$ in the vicinity of $p$:
\begin{eqnarray}
    \begin{aligned}
        &G(\omega) \approx \frac{\alpha(H, p, -\nu_p)}{(\omega - p)^{\nu_p}} \\
        &G'(\omega) \approx \frac{-\nu_p~\alpha(H, p, -\nu_p)}{(\omega - p)^{\nu_p+1}} \\
        &F(\omega) = \frac{G'(\omega)}{G(\omega)} \approx \frac{-\nu_p}{\omega - p}
    \end{aligned}
\end{eqnarray}
$p$ is a singularity of order $1$ of $F$, and its associated residue is its order $\nu_p$.
Similarly, for $\omega$ in the vicinity of a zero $z$ of $G$:
\begin{eqnarray}
    \begin{aligned}
        &G(\omega) \approx \alpha(G, z, -\nu_z)(\omega - z)^{\nu_z} \\
        &G'(\omega) \approx \nu_z~\alpha(G, z, -\nu_z)(\omega - z)^{\nu_z-1} \\
        &F(\omega) = \frac{G'(\omega)}{G(\omega)} \approx \frac{\nu_z}{\omega - z}
    \end{aligned}
\end{eqnarray}
Therefore:
\begin{eqnarray}
    \begin{aligned}
        &\forall (p, \nu_p) \in \mathcal{P}_G, ~ Res(F, p) = -\nu_p \\
        &\forall (z, \nu_z) \in \mathcal{Z}_G, ~ Res(F, z) = +\nu_z
    \end{aligned}
\end{eqnarray}
where $\mathcal{P}_G$ and $\mathcal{Z}_G$ are respectively the sets of poles and zeros of $G$. Knowing the poles and residues of $F$, it is possible to apply MOSEM (Eq.~(\ref{eq:ASE_exp})):
\begin{eqnarray}
    \begin{aligned}
        F(\omega) &= \frac{G'(a)}{G(a)} +\sum_{z_\ell} \left( \frac{\nu_{z_\ell}}{\omega-z_\ell} - \frac{\nu_{z_\ell}}{a-z_\ell} \right) \\
        &- \sum_{p_\ell} \left( \frac{\nu_{p_\ell}}{\omega-p_\ell} - \frac{\nu_{p_\ell}}{a-p_\ell} \right)
    \end{aligned}
    \label{eq:Log_deriv_simple}
\end{eqnarray}
By definition of $F$, $Log(G)$ can be obtained by integrating this expansion on a path from $a$ to $\omega$:
\begin{eqnarray}
    \begin{aligned}
        Log(\frac{G(\omega)}{G(a)}) = &\left(\frac{G'(a)}{G(a)} - \sum_{z_\ell} \frac{\nu_{z_\ell}}{a-z_\ell} + \sum_{p_\ell} \frac{\nu_{p_\ell}}{a-p_\ell} \right) (\omega-a) \\
        &+ \sum_{z_\ell} \nu_{z_\ell}Log(\frac{\omega-z_\ell}{a-z_\ell}) \\
        &- \sum_{p_\ell} \nu_{p_\ell}Log(\frac{\omega-p_\ell}{a-p_\ell}) \\
    \end{aligned}
    \label{eq:Log_deriv_integ}
\end{eqnarray}
$G(a)$ and $G'(a)$ are easily obtained with Eq.~(\ref{eq:weierstrass_fact}) if a is not $0$:
\begin{eqnarray}
    \begin{aligned}
        &G(a) = \frac{H(a)}{a^m} \\
        &G'(a) = \frac{\partial_\omega[H](a)}{a^m} - m\frac{H(a)}{a^{m+1}}
    \end{aligned}
\end{eqnarray}
Additionally, using the Leibniz rule, we can express $G(0)$ and $G'(0)$ from the derivatives of $H$:
\begin{eqnarray}
    \begin{aligned}
        &G(0) = \frac{1}{m!}\partial_\omega^m[H](0) \\
        &G'(0) = \frac{1}{(m+1)!}\partial_\omega^{(m+1)}[H](0)
    \end{aligned}
\end{eqnarray}
Finally, the composition of Eq.~(\ref{eq:Log_deriv_integ}) with the exponential function is performed to retrieve the general Singularity and Zero Factorization (SZF) of the meromorphic function $H$:
\begin{eqnarray}
    \begin{aligned}
        &H(\omega) = \omega^m ~ G(a) ~ \frac{\prod_{z_\ell}(1 - \frac{\omega-a}{z_\ell-a})^{\nu_{z_\ell}}}{\prod_{p_\ell}(1 - \frac{\omega-a}{p_\ell-a})^{\nu_{p_\ell}}} ~ e^{i\tau(\omega-a)} \\
        &\tau = -i\left(\frac{G'(a)}{G(a)} - \sum_{z_\ell} \frac{\nu_{z_\ell}}{a-z_\ell} + \sum_{p_\ell} \frac{\nu_{p_\ell}}{a-p_\ell} \right)
    \end{aligned}
    \label{eq:Log_deriv_exp}
\end{eqnarray}
The SZF is similar to the pole and zero factorization obtained in the case of electronics or automatics where the transfer function is a ratio of complex polynomial~\cite{sanathanan1963}, or the ones obtained in reference~\cite{grigoriev2013} in the case of simple poles.

\subsection{Constraints on the Zeros and the Time Constant}

Let $K$ be the meromorphic function defined, as $K(\omega) = H(-\omega^*)^*$. Since we consider a real signal $h$, we must have the hermitian symmetry $K=H$. Therefore, if $z$ is a non-null zero of $H$, then $K(z)=H(-z^*)=0$, \textit{i.e.} $-z^*$ is also a zero, and the SZF shows that its order is $\nu_z$. Therefore:
\begin{eqnarray}
    \begin{aligned}
        H(z)&=0 \Rightarrow H(-z^*)=0 \\
        &\nu_{-z^*} = \nu_z
    \end{aligned}
\end{eqnarray}
Additionally, we show that the time constant $\tau=\tau_R + i\tau_I$ must be real, with
\begin{eqnarray}
    \begin{aligned}
    \tau_R = &Im\left[\frac{G'(a)}{G(a)}\right] - \sum_{z_\ell} Im\left[\frac{\nu_{z_\ell}}{a-z_\ell}\right] \\
    &+ \sum_{p_\ell} Im\left[\frac{\nu_{p_\ell}}{a-p_\ell}\right] \\
    \tau_I = -&Re\left[\frac{G'(a)}{G(a)}\right] + \sum_{z_\ell} Re\left[\frac{\nu_{z_\ell}}{a-z_\ell}\right] \\
    &- \sum_{p_\ell} Re\left[\frac{\nu_{p_\ell}}{a-p_\ell}\right]
    \end{aligned}
\end{eqnarray}
In the case where $a=0$, this simplifies into:
\begin{eqnarray}
    \begin{aligned}
        \tau_R = &Im\left[\frac{G'(0)}{G(0)}\right] - 2\text{i}\sum_{p_\ell\in\text{i}\mathbb{R}} \frac{\nu_{p_\ell}}{p_\ell}
    \end{aligned}
\end{eqnarray}
and
\begin{eqnarray}
    \begin{aligned}
        \tau_I = -&Re\left[\frac{G'(0)}{G(0)}\right]
    \end{aligned}
\end{eqnarray}
Let us consider the real frequency $\omega$. The hermitian symmetry leads to:
\begin{eqnarray}
    \begin{aligned}
        ~\omega^m~e^{i\tau_R\omega}~e^{-\tau_I\omega} = (-1)^m~\omega^m~e^{i\tau_R\omega}~e^{\tau_I\omega}
    \end{aligned}
\end{eqnarray}
Which is not satisfied as $\omega$ tends towards $+\infty$ unless $\tau_I=0$. Taking into accounts the constraints on the poles, and for $a=0$ this leads to:
\begin{eqnarray}
    \begin{aligned}
        \tau = &-\text{i} \left(\frac{G'(0)}{G(0)} - \sum_{p_\ell\in\text{i}\mathbb{R}} \frac{\nu_{p_\ell}}{p_\ell} \right) \\
        &+ 2Im\left(\sum_{z_\ell, Re[z_\ell]>0} \frac{\nu_{z_\ell}}{z_\ell} - \sum_{p_\ell, Re[p_\ell]>0} \frac{\nu_{p_\ell}}{p_\ell} \right) 
    \end{aligned}
\end{eqnarray}
with $G'(0)/G(0) \in \text{i}\mathbb{R}$.

\subsection{Accuracy of the SZF}

Similarly to the previous section, we show the accuracy of the SZF on the function $H$ defined in Eq.~(\ref{eq:test_function_fact}). The poles were already calculated: 
\begin{eqnarray}
    \begin{aligned}
        &\omega_p^{(n)} &= \frac{(2n+1)\pi}{2}, ~ n\in\mathbb{N}
        \end{aligned}
\end{eqnarray}
The only zero is $1$, and the derivation of $H(0)=-1$ and $H'(0)=-1\text{i}$ is straightforward. If we consider $M$ poles, we obtain the following approximation of $\tau$:
\begin{eqnarray}
    \begin{aligned}
        &\tau_M = -\text{i}\left(\text{i} + 1 - \sum_{n=-M}^{M} \frac{2}{\omega_p^{(n)}} \right)
    \end{aligned}
\end{eqnarray}
Finally, we write the truncated SZF of $H$:
\begin{eqnarray}
    \begin{aligned}
        H(\omega) &\approx e^{\text{i}\tau_M\omega} \frac{1-\omega}{\prod_{n=-M}^{M} (1 - \frac{\omega}{\omega_p^{(n)}})^2}
    \end{aligned}
    \label{eq:szf_test}
\end{eqnarray}

\begin{figure}
    \centering
    \includegraphics[width=0.9\columnwidth]{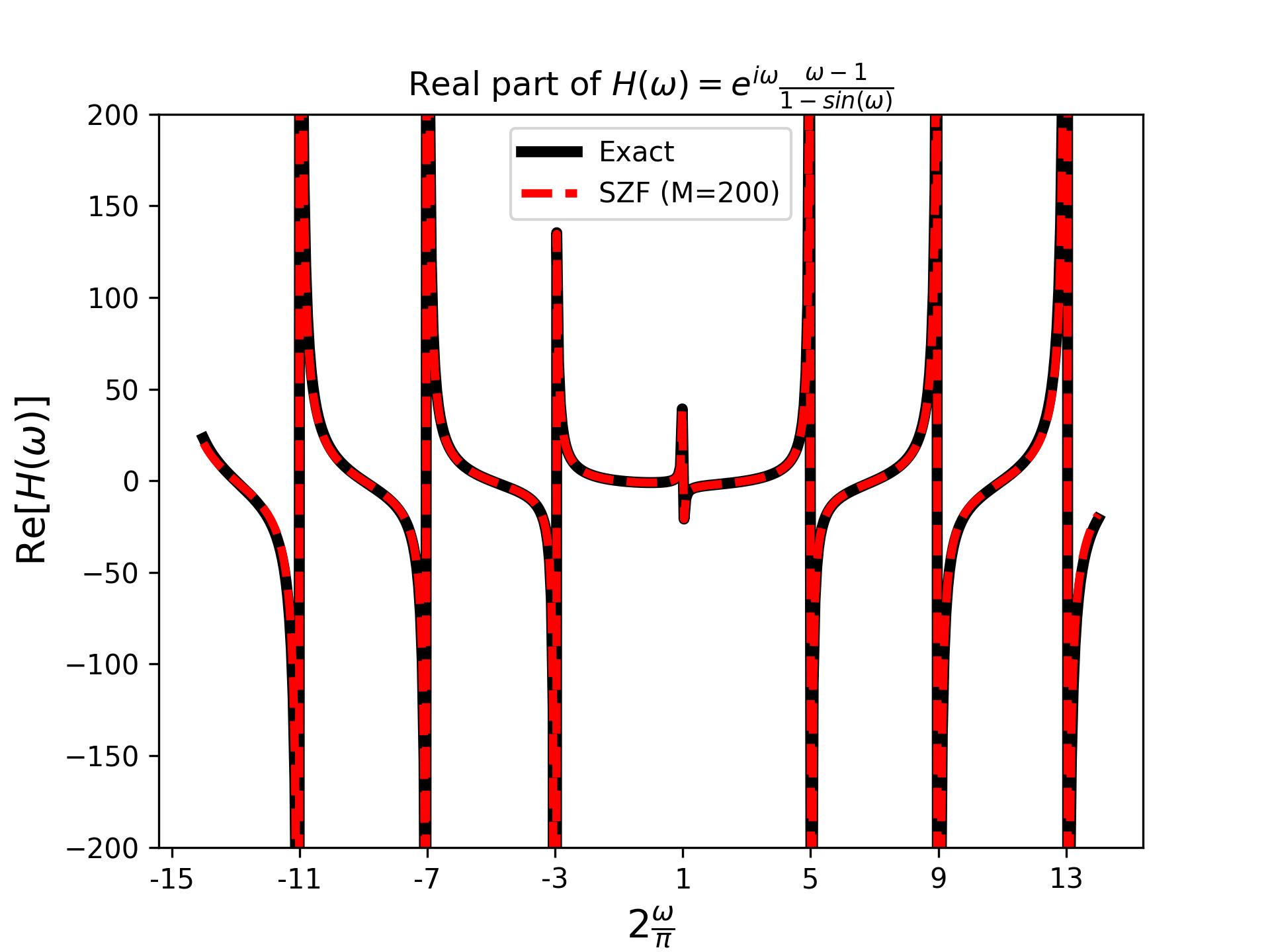}
    \caption{Real part of the function $H$ defined in Eq.~(\ref{eq:test_function_fact}). The poles of $H$ are the complex frequencies $\omega_p^{(n)}=2n\pi + \pi/2$. The exact expression in black is compared to the truncated SZF in red: $-M \leq n \leq M$, with $M=200$, given in Eq.~(\ref{eq:szf_test}).}
    \label{fig:order2_harmonic_szf}
\end{figure}
We compare the truncated SZF with $M=200$ poles to the exact expression of $H$ in Fig.~\ref{fig:order2_harmonic_szf}. Let us point out that many more poles $M$ are usually necessary for the SZF than for MOSEM to highly reduce the error between the fitted function and its approximation.

\section{Temporal Expression with MOSEM}\label{sec:temporal_exp}

\subsection{Inverse Laplace Transform with MOSEM}
The inverse Laplace transforms of the base functions $F_{m,p}$ are calculated. These base functions are defined in the harmonic domain as:
\begin{eqnarray}
    \begin{aligned}
        &F_{m,p}(\omega) = \frac{1}{(\omega - p)^m} \\
        &f_{m,p}(\omega) = \mathcal{L}^{-1}[F_{m,p}](\omega)
    \end{aligned}
\end{eqnarray}
By recurrence, the inverse Laplace transform $f_{m,p}$ of $F_{m,p}$ can be expressed from $f_{1,p}$:
\begin{eqnarray}
    f_{n,p}(t) = \frac{(-it)^{(n-1)}}{(n-1)!}f_{1,p}(t)
    \label{eq:Recc_FN}
\end{eqnarray}
In addition:
\begin{eqnarray}
    f_{1,p}(t) = -ie^{-ip t} u(t)
    \label{eq:f1}
\end{eqnarray}
This leads to the following expression of the dynamic states $f_{n,p}$:
\begin{eqnarray}
    f_{n,p}(t) = \frac{(-i)^n}{(n-1)!}t^{n-1}e^{-ipt}u(t)
    \label{eq:fn}
\end{eqnarray}
The inverse Laplace transform $h$ of the harmonic-domain function $H$ can therefore be derived the singularity expansion obtained with MOSEM (Eq.~(\ref{eq:ASE_exp}):
\begin{eqnarray}
    \label{eq:transfer_temporal}
    \begin{aligned}
        h(t) &= H_{\text{NR}}\delta(t) \\
        &+ \sum_{p} \sum_{m=1}^{\nu_p} \alpha(H, p, -m) f_{m, p}(t)
    \end{aligned}
\end{eqnarray}

\bibliography{apssamp}
\bibliographystyle{iopart-num.bst}